\documentstyle[multicol,aps,epsfig,amsbsy]{revtex}

\newcommand{\PkPop}    { {P_\Pi} }
\newcommand{\BCTime}   { { \langle t \rangle } }
\newcommand{\AvFit}    { { \langle f \rangle } }
\newcommand{\MuOp}     { {\rm M} }

\begin{document}

\title{Metastable Evolutionary Dynamics:\\
Crossing Fitness Barriers or\\
Escaping via Neutral Paths?}
\author{Erik van Nimwegen\thanks{Permanent address: Bioinformatics
Group, University of Utrecht, Padualaan 8, NL-3584-CH Utrecht,
The Netherlands}
and James P. Crutchfield}
\address{Santa Fe Institute, 1399 Hyde Park Road, Santa Fe, NM 87501\\
Electronic addresses: \{erik,chaos\}@santafe.edu}

\date{\today}
\maketitle
\begin{abstract}
We analytically study the dynamics of evolving populations that exhibit
metastability on the level of phenotype or fitness. In constant
selective environments, such metastable behavior is caused by two
qualitatively different mechanisms. One the one hand, populations may
become pinned at a local fitness optimum, being separated from
higher-fitness genotypes by a {\em fitness barrier} of low-fitness
genotypes. On the other hand, the population may only be metastable on
the level of phenotype or fitness while, at the same time, diffusing
over {\em neutral networks} of selectively neutral genotypes.
Metastability occurs in this case because the population is separated
from higher-fitness genotypes by an {\em entropy barrier}: The
population must explore large portions of these neutral networks
before it discovers a rare connection to fitter phenotypes.

We derive analytical expressions for the barrier crossing times in
both the fitness barrier and entropy barrier regime. In contrast with
``landscape'' evolutionary models, we show that the waiting times to
reach higher fitness depend strongly on the width of a fitness
barrier and much less on its height. The analysis further shows that
crossing entropy barriers is faster by orders of magnitude than
fitness barrier crossing. Thus, when populations are trapped in a
metastable phenotypic state, they are most likely to escape by
crossing an entropy barrier, along a neutral path in genotype
space. If no such escape route along a neutral path exists, a
population is most likely to cross a fitness barrier where the barrier
is {\em narrowest}, rather than where the barrier is shallowest.

\begin{center}
Santa Fe Institute Working Paper 99-07-041
fitness barrier, entropy barrier, metastability.\\
{\em Running Head}: Metastable Evolutionary Dynamics
\end{center}
\end{abstract}


\begin{multicols}{2}

\vspace{-0.25in}
\tableofcontents

\section{Introduction}

For populations evolving under selection, mutation, and a static
fitness function, there are two main mechanisms thought to be
responsible for the occurrence of dynamical metastability---a
behavior commonly observed in natural and artificial evolutionary
processes
\cite{Adam95a,Barnett98,Crutchfield&Mitchell95a,Elena&Cooper&Lenski96,Font98a,Newman&Engelhardt98}
and called {\em punctuated equilibria} in paleobiology
\cite{Gould&Eldredge77}. First, a population may become trapped
around a local optimum in the fitness ``landscape'' until a rare
mutant crosses a {\em fitness barrier} to a higher nearby
peak. Second, more recently it has been proposed
\cite{Font98a,Huynen&Stadler&Fontana,Nimw97a} that populations may
evolve neutrally, drifting randomly over {\em neutral networks} of
isofitness genotypes in genotype space, until a rare single-point
mutant connection is found to another neutral network of higher fitness.
In this case, the population must cross an {\em entropy barrier} by
visiting a large volume of the neutral network before it discovers a
path to higher fitness.

To understand the relative roles of these two mechanisms in
evolutionary metastability, in the following we study the
dynamics of a population evolving under simple fitness functions that
contain a single fitness barrier of tunable height and width. In order
for the population to escape its current metastable state and so reach
higher fitness, it must create a genotype that is separated from the
current fittest genotypes in the population by a {\em valley}
of lower-fitness genotypes. The {\em height} of the fitness barrier
measures the relative selective difference between the current fittest
genotypes and the lower-fitness genotypes in the intervening
valley. Its {\em width} denotes the number of point mutations the
current fittest genotypes must undergo to cross the valley of
low fitness genotypes. We derive explicit analytical predictions for
the barrier crossing times as a function of population size, mutation
rate, and barrier height and width. The scaling of the fitness-barrier
crossing time as a function of these parameters shows that the waiting
time to reach higher fitness depends crucially on the width of the
barrier and much less on the barrier height.

This contrasts with the scaling of the barrier crossing time for a
particle diffusing in a double-well potential---a model proposed
previously for populations crossing a fitness barrier
\cite{Lande85,Newman&Cohen&Kipnis85}. For such stochastic processes,
it is well known that the waiting time scales exponentially with the
barrier height \cite{Gardiner85}. In the population dynamics that we
analyze here, we find that the waiting time scales approximately
exponential with barrier width and only as a power law of the
logarithm of barrier height. In addition, the waiting time scales
roughly as a power law in both population size and mutation rate.

When the barrier height is lowered below a critical height, the
fitness barrier turns into an entropy barrier. We show that, in
general, neutral evolution via crossing entropy barriers is faster
by orders of magnitude than fitness barrier crossing. Additionally,
we show that the waiting time for crossing entropy barriers exhibits
anomalous scaling with population size and mutation rate.

Finally, we extend our analysis to a class of more complicated fitness
functions that contain a network of tunable fitness and entropy
barriers. We show that the theory still accurately predicts fitness-
and entropy-barrier crossing times in these more complicated cases.

The general conclusion drawn from our analysis is that, when
populations are trapped in a metastable phenotypic state, they are most
likely to escape this metastability by crossing an entropy barrier.
That is, the escape to a new phenotype occurs along a neutral path in
genotype space. If no such neutral path exists, then the population is
most likely to cross a fitness barrier at the place where the barrier is
{\em narrowest}.

\subsection{Evolutionary Pathways and Metastability}

The notion of an {\em adaptive landscape}, first introduced by Wright
\cite{Wright32}, has had a large impact on our appreciation of the
mechanisms that control how populations evolve in static environments.
The intuitive idea is that a population moves up the slopes of its
fitness ``landscape'' just as a physical system moves down the slope
of its potential-energy surface. Once this analogy has been accepted,
it is natural to borrow many of the qualitative results on the dynamics
of physical systems to account for the dynamics of evolving populations.
For instance, it has become common to assume that an evolving population
can be modeled by a single uphill walker in a ``rugged'' fitness
landscape \cite{Kauf87a,Mack89a}. 

There are, however, seemingly different kinds of evolutionary behavior
than incremental adaptation via ``landscape'' crawling. For example,
metastability or punctuated equilibrium of phenotypic traits in an
evolving population appears to be a common occurrence in biological
evolution \cite{Elena&Cooper&Lenski96,Gould&Eldredge77} as well as in
models of natural and artificial evolution
\cite{Adam95a,Crutchfield&Mitchell95a,Font98a}. As just pointed out,
for simple cases where populations evolve in a relatively constant
environment, there are two main mechanisms that have been proposed to
account for this type of metastable behavior.

The first and most commonly accepted explanation was already implicit
in Wright's {\em shifting balance} theory \cite{Wright82a}. A
population moves up the slope of its fitness ``landscape'' until it
reaches a local optimum, where it stabilizes. The population is
pictured as a cloud in genotype space focused around this local
optimum.  The population remains in this state until a rare sequence
of mutants crosses a {\em valley} of low fitness towards a higher
fitness peak.  In this view, metastability is the result of {\em
fitness barriers} that separate local optima in genotype space.

This mechanism for metastability is very reminiscent of that found in
physical systems. Metastability occurs there because local energy
minima in state space are separated by potential energy barriers,
which impede the immediate transition between the minima. A physical
system generally moves through its state space along trajectories that
lower its energy. Once it reaches a local minimum it tends to stay
there. However, when such a system is subject to thermal fluctuations,
through a sequence of chance events it can eventually be pushed over a
barrier that separates the current local minimum from another. When
this transition occurs, it turns out that the system moves quickly
to the new local minimum.

Mathematically, barrier crossing processes in physical systems are most
often described as diffusion in a potential field, where the potential
represents the energy ``landscape''. These processes have been
extensively studied and the basic quantitative results are widely known
\cite{Frau97a,Gardiner85,vanKampen92}. For example, barrier crossing
times increase exponentially with the {\em height} of the barrier and
inverse exponentially with the fluctuation amplitude, as measured by
temperature.

In light of the physical metaphor for evolving populations, it is not
surprising that the dynamics of populations crossing fitness barriers
has been modeled using a class of diffusion equations analogous to
those used to describe thermally driven systems in a potential
\cite{Lande85,Newman&Cohen&Kipnis85}. In this approach, the dynamics
of the average fitness of the population is modeled as diffusion over
the ``fitness landscape'', thermal fluctuations are replaced by random
genetic mutations and drift, and the population size, which controls
sampling stochasticity, plays the role of inverse temperature. As a
direct consequence, it was found that fitness-barrier crossing times
scale exponentially with population size in these models. Note that it
is assumed in this approach that the population {\em as a whole} must
cross the fitness barrier.

In the following, we show that the analogy with the physical situation
and, in particular, the translation of results from there are
misleading for the understanding of the evolutionary dynamics. For
example, a direct analysis of the population dynamics reveals that for
most parameter ranges, the time to cross a fitness barrier scales very
differently for populations evolving under selection and mutation. For
example, the waiting time is determined by how long it takes to
generate a {\em rare sequence of mutants} that crosses the fitness
barrier, as opposed to how long it takes the population {\em as a whole}
to cross the fitness barrier.

This brings us to the second main mechanism for metastability---one
that has been put forward more recently
\cite{Barnett98,Font98a,Huynen&Stadler&Fontana,Newman&Engelhardt98,Nimw97a}.
The second mechanism derives from the observation that large sets of
mutually fitness-neutral genotypes are interconnected along single-point
mutation paths. That is, sets of isofitness genotypes form extended
{\em neutral networks} under single-point mutations in genotype space.

In this alternative scenario, a population displays a constant
distribution of phenotypes for some period while, at the same time,
individuals in the population diffuse over a neutral network in
genotype space. That is, despite phenotypic metastability, there is no {\em
genotypic} stasis during this period. The phenotype distribution
remains metastable until, via diffusion over the neutral network, a
member of the population discovers a genotypic connection to a
higher-fitness neutral network.

When this mechanism operates, metastability is the result of an {\em
entropy barrier}, as we call it. The population must spread over or
search large parts of the neutral network before it finds a connection
to a higher-fitness network. One envisages the population moving
randomly through a genotypic labyrinth of common phenotypes with only
a single or relatively few exits to fitter phenotypes.

\subsection{Overview}

In the following, we analyze and compare the population dynamics of
crossing such fitness and entropy barriers with the goals of
elucidating the basic mechanisms responsible for each, calculating
the scaling forms for the evolutionary times scales associated with
each, and understanding their relative importance when
both can operate simultaneously.

Section \ref{evolutionary_dynamics} defines the basic evolutionary
model.

Section \ref{crossing_single_barrier} introduces a tunable
fitness function that models the simplest case in which to study both
types of barrier crossing. It consists of a single local optimum, with a
valley, and a single portal (target genotype) in genotype space. By
tuning the height of the local optimum one can change the fitness
barrier into an entropy barrier. We analyze this basic model as a
branching process, calculating the statistics of lineages of
individuals in the fitness valley. Comparison of the theoretical
predictions for the fitness-barrier crossing times with data obtained
from simulations shows that the theory accurately predicts
these fitness-barrier crossing times for a wide range of
parameters. We also derive several simple scaling relations for the
fitness-barrier crossing times appropriate to different parameter
regimes.

Section \ref{err_thres_and_neut_reg} first determines the barrier
heights at which the fitness-barrier regime shifts over into an
entropic one.  After this, we discuss the population dynamics of
crossing entropy barriers, providing rough scaling relations for the
barrier crossing times in this regime. Comparison of these results
with the scaling relations for fitness-barrier crossing shows that
entropy-barrier crossing proceeds markedly faster than crossing
fitness barriers.

Section \ref{royal_staircase} extends our analysis to a set of much more
complicated fitness functions---a class called the {\em Royal Staircase
with Ditches}. These fitness functions are closely related to the
Royal Road \cite{Nimw97a,Nimw97b} and Royal Staircase
\cite{Nimw98b,Nimw98a} fitness functions that we studied earlier, which
consist of a sequence or a network of entropy barriers only. The Royal
Staircase with Ditches generalizes this class of fitness functions to
one that possesses multiple fitness and entropy barriers of variable
width, height, and volume. We adapt the theoretical analysis using our
statistical dynamics approach \cite{Nimw97b} to deal with these more
complicated, but more realistic cases. Comparison of the theoretically
predicted and experimentally obtained barrier crossing times again
shows that the theory accurately predicts the barrier crossing times
in these more complicated situations as well. 

Finally, Sec. \ref{conclusions} presents our conclusions and discusses
the general picture of metastable population dynamics that emerges
from our analyses.

\section{Evolutionary Dynamics}
\label{evolutionary_dynamics}

We consider a simple evolutionary dynamics of selection
and mutation with a constant population size $M$. An individual's
genotype consists of a binary sequence of on-or-off genes. We
consider the simple case in which the fitness of an individual is
determined by its genotype only. The genotype-to-phenotype and
phenotype-to-fitness maps are collapsed into a direct determination
of a genotype's fitness. Selection and reproduction are assumed to
take place in discrete generations, with mutation occurring at
reproduction. The exact evolutionary dynamics is defined as follows.
\begin{itemize}
\item{A {\em population} consists of $M$ binary sequences of a fixed
size $L$.}
\item{A {\em fitness} $f_s$ is associated with each of the $2^L$
possible genotypes $s \in {\cal A}^L$, where ${\cal A} = \{0,1\}$.}
\item{Every generation $M$ individuals in the current population are
sampled with replacement and with a probability proportional to their
fitness. Thus, the expected number of offspring for an individual with
genotype $s$ is $f_s/\AvFit$, where $\AvFit$ is
the current average fitness of the population.}
\item{Once the $M$ individuals have been selected, each bit in each
individual is mutated (flipped) with probability $\mu$, the {\em
mutation rate}.}
\end{itemize}
In this basic model there are effectively two evolutionary parameters:
the mutation rate $\mu$ and the population size $M$. 

Several aspects of the basic model---such as, discrete generations and
fixed population size---were mainly chosen for analytical convenience.
The discrete-generation assumption can be lifted, leading to a
continuous-time model, without affecting the results presented below.
As for the assumption of fixed population size, the analysis can be
adapted in a straightforward manner to address (say) fluctuating or
exponentially growing populations.

Models including genetic recombination are notoriously more difficult
to analyze mathematically. Despite our interest in the effects of
recombination, it is not included here largely for this reason.
Moreover, for wide parameter ranges in the neutral and
piecewise-neutral evolutionary processes we consider, it appears that
recombination need not be a dominant mechanism. For example, Refs.
\cite[sec. 6.5]{Nimw97b} and \cite[sec. VIII]{Nimw98a} show that
recombination often does not significantly affect population dynamics
in these cases.

\section{Crossing a Single Barrier}
\label{crossing_single_barrier}

We first consider the simple case of a single barrier for the
population to cross. Of the $2^L$ genotypes, there is one with fitness
$\sigma > 1$ that we refer to as the {\em peak} genotype $\Pi$. Then
there are $2^L-2$ genotypes with fitness $1$ that we refer to as {\em
valley} genotypes. Finally, there is a single {\em portal} genotype
$\Omega$ at a Hamming distance $w$ from the fitness-$\sigma$ peak
genotype $\Pi$. We view the portal genotype as giving access to
higher-fitness genotypes---genotypes whose details are unimportant,
since in this section we only analyze the dynamics up to the portal's
first discovery.

\begin{figure}[htbp]
\centerline{\epsfig{file=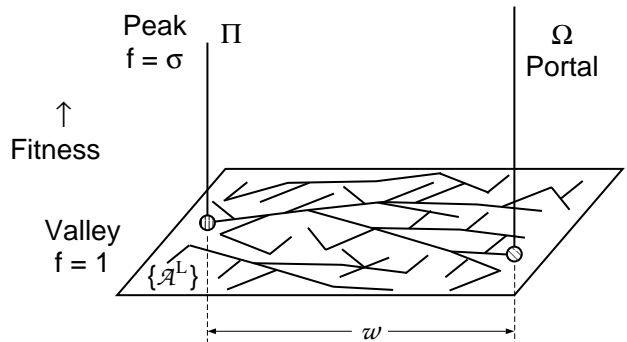,width=3.25in}}
\caption{Evolution from the peak genotype $\Pi$ to the higher-fitness
portal genotype $\Omega$ via low-fitness valley genotypes. The selective
  advantage of the peak individuals over those in the valley is
  controlled by the peak height $\sigma$. The portal and peak genotypes
  are a Hamming (mutational) distance $w$ apart. The domain is the
  hypercube ${\cal A}^L$ of all length-$L$ genotypes.}
\label{PeakValley}
\end{figure}

The variable $w$ tunes the fitness barrier's {\em width}
and the variable $\sigma$ its {\em height}. The height $\sigma$ also
indicates a peak individual's {\em selective advantage} over those in
the valley, as measured by the relative difference $\sigma-1$ of
their expected number of offspring. Figure \ref{PeakValley}
illustrates the basic setup.

At time $t=0$ the population starts with all $M$ genotypes located at
the peak $\Pi$. We then evolve the population under selection and
mutation, as described in the previous section, until the portal
genotype $\Omega$ occurs
in the population for the first time. (Hence, the portal's fitness is
not relevant.) This defines one evolutionary {\em run}. We record the
time $t$ at which the portal is discovered. We are interested in the
average discovery time $\langle t \rangle$, averaged over an ensemble
of such runs. We are particularly interested in the scaling of this
{\em barrier crossing time} $\langle t \rangle$ as a function of the
evolutionary parameters $M$ and $\mu$, as well as the barrier
parameters $\sigma$ and $w$.

Let's briefly review in simple language the evolutionary dynamics before
launching into the mathematical analysis. In the parameter regime with
$\sigma \gg 1$, where the peak fitness is considerably larger than the
valley fitness, and with the mutation rate $\mu$ not too high, the bulk
of the population remains at the peak. That is, the population is a
{\em quasispecies} cloud, centered around the peak genotype $\Pi$
\cite{Eigen&McCaskill&Schuster89}. For such parameter regimes, the
barrier is clearly a fitness barrier: the waiting time $\langle t
\rangle$ is determined by the time it takes to create a rare sequence
of mutant genotypes that crosses the valley between the peak and the
portal.

However, as $\sigma \rightarrow 1^+$, the fitness barrier transforms
into an entropy barrier. For $\sigma =1$ there is no fitness difference
between peak and valley genotypes and the entire population simply
diffuses through genotype space until the portal is discovered. As we
will see below, the entropic regime sets in rather suddenly at a value
of $\sigma_c$ somewhat above $\sigma =1$. As we show, this transition
is the well known {\em error threshold} of molecular evolution theory
\cite{Eigen71}. At $\sigma = \sigma_c$, the value of which depends on
the population size $M$ and mutation rate $\mu$, the subpopulation on
the peak becomes unstable in the sense that all individuals on the
peak may be lost through a fluctuation. More precisely, the waiting
time for such a fluctuation to occur becomes short in comparison to
the fitness-barrier crossing time. When this fluctuation has occurred,
there is no longer a restoring ``force'' that keeps the population
concentrated around the peak genotype. The population as a whole
diffuses randomly through the valley as if the genotypes were all
fitness neutral. While our analysis accurately predicts the barrier
crossing times in the fitness-barrier regime, it is notable that beyond
the error threshold, in the entropic regime, only order-of-magnitude
predictions can be obtained using the current analytical tools.

Calculating the barrier crossing time proceeds in three stages. First,
in Sec. \ref{metastable_quasispecies} we determine the population's
quasispecies distribution, defined as the average proportions of
individuals located on the peak and in the valley during the metastable
state. From this, one directly calculates the average fitness in the
population. Second, in Sec. \ref{valley_lineages} we consider the
genealogy statistics of individuals in the valley. In the
fitness-barrier regime, genealogies in the valley are generally
short-lived and are all seeded by mutants of the peak genotype. We
approximate the evolution of valley genealogies as a branching process
and use this representation to calculate the average barrier crossing
time. Third, with this analysis complete, Sec.
\ref{crossing_fit_barrier} then addresses the transition from the
fitness-barrier regime to the entropic one.

\subsection{Metastable Quasispecies}
\label{metastable_quasispecies}

Each evolutionary run, the population starts out concentrated at the
peak genotype $\Pi$. After a relaxation phase, assumed to be short
compared to the barrier crossing time, there will be roughly constant
proportions of the population on the peak and in the valley. We now
calculate the equilibrium proportion $\PkPop$ of peak individuals and
the population's average fitness $\AvFit$, after this
relaxation phase.

To first approximation, one can neglect {\em back mutations} from valley
individuals back into the peak genotype. First of all, if $\sigma\gg 1$,
selection keeps the bulk of the population on the peak. Additionally,
valley individuals produce fewer offspring than peak individuals and
they are unlikely---with a probability $1/L$ at most---to move back
onto the peak when they mutate. In this regime, the quasispecies
distribution is largely the result of a balance between selection
expanding the peak population by a factor of $\sigma/\AvFit$
and deleterious mutations moving them into the valley with probability
$1-(1-\mu)^L$. The result is that we have a balance equation for the
proportion $\PkPop$ of peak individuals given by
\begin{equation}
\PkPop = \frac{\sigma}{\AvFit} (1-\mu)^L \PkPop ~.
\label{BalanceEquation}
\end{equation}
From this we immediately have that
\begin{equation}
\AvFit = \sigma (1-\mu)^L ~.
\label{av_fit_approx}
\end{equation}
Since we also have that
$\langle f \rangle = \sigma P_{\Pi} + 1 \cdot (1-P_{\Pi})$, we can
determine the proportion of peak individuals to be:
\begin{equation}
\label{peak_prop_approx}
\PkPop = \frac{\AvFit -1}{\sigma -1} ~.
\label{PkIndividuals}
\end{equation}
For parameters where $\AvFit = \sigma (1-\mu)^L \gg 1$,
Eqs. (\ref{av_fit_approx}) and (\ref{peak_prop_approx}) give quite
accurate predictions for the average fitness and the proportion of
individuals on the peak.

In cases where $\AvFit$ is close to $1$, a substantial
proportion of the population is located in the valley and back
mutations from the valley onto the peak must be taken into account. To
do this, we introduce the quasispecies Hamming distance distribution
$\vec{P} = ( P_0, \ldots, P_i, \ldots, P_L )$, where $P_i$ is the
proportion of individuals located at Hamming distance $i$ from $\Pi$.
Thus, $P_0 = \PkPop$ indicates the proportion on the peak. Under
selection, the distribution $\vec{P}$ changes according to:
\begin{equation}
\label{selec_type_i}
P^{{\rm sel}}_i =
  \frac{(\sigma-1) \delta_{i0} + 1}{\AvFit} P_i ~,
\end{equation}
where $\delta_{ij} = 1$, if $i=j$, and $\delta_{ij} = 0$, otherwise.
We can write this formally as the result of an operator acting on
$\vec{P}$:
\begin{equation}
\vec{P}^{{\rm sel}} = \frac{\left({\bf S} \cdot \vec{P}\right)_i}{\AvFit} ~,
\end{equation}
where 
\begin{equation}
S_{ij} = \left[ (\sigma-1) \delta_{i0} + 1 \right] \delta_{ij} ~,
\end{equation}
defines the {\em selection operator} $\bf S$.

Next, we consider the transition probabilities $\MuOp_{ij}$ that under
mutation a genotype at Hamming distance $j$ from the peak moves to
a genotype at Hamming distance $i$ from the peak. We have that:
\begin{eqnarray}
\nonumber
\MuOp_{ij} & = & \sum_{u=0}^{L-j} \sum_{d=0}^j \delta_{j+u-d,i} {L-j
\choose u} {j \choose d} \\
  && \times \mu^{u+d} (1-\mu)^{L-u-d} ~.
\label{MutationOperator}
\end{eqnarray}
That is, $\MuOp_{ij}$ is the sum of the probabilities of all possible ways
to mutate $u$ of the $L-j$ bits shared with $\Pi$ and $d$ of the $j$
bits that differ, such that $j+u-d=i$. Equation (\ref{MutationOperator})
defines the {\em mutation operator} $\bf M$.

We can now introduce the {\em generation operator} ${\bf G} = {\bf M}
\cdot {\bf S}$. The equilibrium quasispecies distribution $\vec{P}$ is
a solution of the equation
\begin{equation}
\vec{P} = \frac{{\bf G} \cdot \vec{P}} {\AvFit} ~.
\end{equation}
In this way, the quasispecies distribution is given by the principal
eigenvector, normalized in probability, of the matrix ${\bf G}$; while
the average fitness $\AvFit$ is given by ${\bf G}$'s principal
eigenvalue. Note that this is conventional quasispecies theory
\cite{Eigen&McCaskill&Schuster89}, apart from the facts that we have
grouped the quasispecies members into Hamming-distance classes and
that we consider discrete generations, rather than continuous time.

\subsection{Valley Lineages}
\label{valley_lineages}

Under the approximation that back mutations from the valley onto the
peak can be neglected, a roughly constant proportion $1-\PkPop$ of
valley individuals is maintained by a roughly constant influx of
mutants from the peak.  Every generation, some peak individuals leave
mutant offspring in the valley. Additionally, each valley individual
leaves on average a fraction $1/\AvFit$ offspring in the next
generation, as can be seen from Eq. (\ref{selec_type_i}). This means
that the fraction of valley individuals, for which {\em all} of its
$t$ ancestors in the previous $t$ generations were valley individuals
as well, is only $1/\AvFit^t$. For $\AvFit \gg 1$, this implies in
turn that whenever a peak individual seeds a new lineage of valley
individuals by leaving a mutant offspring in the valley, this lineage
is unlikely to persist for a large number of generations. In other
words, lineages composed of valley individuals are short lived.

\begin{figure}[htbp]
\centerline{\epsfig{file=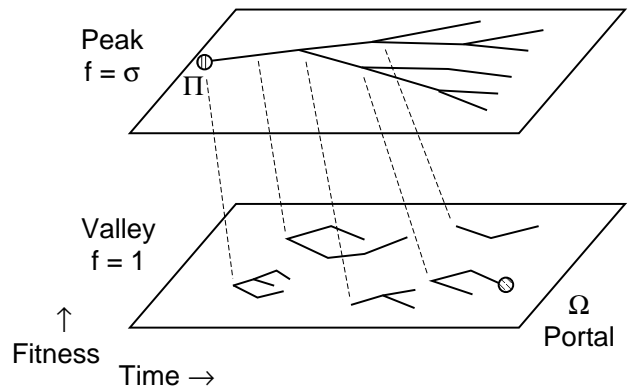,width=3.25in}}
\caption{Genealogies during fitness-barrier crossing. An example
  genealogical tree is sketched for peak individuals (above); they
  have fitness $\sigma$ and are copies of the peak genotype $\Pi$.
  The valley individuals (below) at lower fitness occur in genealogies
  that are seeded (dashed lines) from peak individuals. These
  genealogies are relatively short-lived bushes. Evolution continues
  until the time at which one of the valley bushes discovers the
  portal genotype $\Omega$.}
\label{Genealogies}
\end{figure}

Intuitively, the idea is that the preferred selection of peak
individuals leads to a ``surplus'' of peak offspring that spills into
the valley through mutations. Each mutant offspring of a peak
individual forms the root of a relatively small, i.e., short-lived,
genealogical tree of valley individuals. The barrier crossing time is
determined essentially by the waiting time until one of these
{\em genealogical bushes} produces a descendant that discovers the
portal. These processes are illustrated in Fig. \ref{Genealogies}.

We will analyze the evolution of a valley lineage as a branching
process \cite{Harrisbook}. The probability $O_n$ that one particular
valley individual leaves $n$ offspring in the next generation is given
by a binomial distribution. This is well approximated by a Poisson
distribution as follows:
\begin{eqnarray}
\label{poiss_offspring}
\nonumber
O_n & = & {M \choose n} \left(\frac{1}{M \AvFit} \right)^n
  \left( 1-\frac{1}{M \AvFit} \right)^{M-n} \\
  & \approx & \frac{1}{n!} \left( \frac{1}{\AvFit}\right)^n
  e^{-1 / \AvFit} ~.
\end{eqnarray}
To a good approximation, we may treat the evolution of each valley
lineage as independent of the other valley lineages. Under this
approximation, each valley individual independently has a distribution
of offspring given by Eq. (\ref{poiss_offspring}). Of course, under
fixed population size, the independence assumption may break down when
valley lineages dominate the population.

We now calculate the probability that a valley lineage produces a
descendant that discovers the portal $\Omega$ before the lineage goes
extinct. Let a valley lineage be founded by an ancestor in the valley
that is located at Hamming distance $j$ from $\Omega$. We denote by
$\bar{p}_j(t)$ the probability that $t$ generations from now, {\em
none} of this ancestor's descendants will have discovered the
portal. This probability can be determined recursively, in terms of
the probabilities $\bar{p}_i(t-1)$ as follows,
\begin{eqnarray}
\label{PrDiscoverPortal}
\bar{p}_j(t) & = & O_0 + O_1 \sum_{i=1}^L \bar{p}_i(t-1) \MuOp_{ij} \\
\nonumber
  & + & O_2 \sum_{i,k=1}^L \bar{p}_i(t-1) \MuOp_{ij} \bar{p}_k(t-1) \MuOp_{kj}
  + \ldots ~.
\end{eqnarray}

The first term in the above equation corresponds to the ancestor
having no offspring. This, of course, ensures that the portal will not
be discovered $t$ generations from now, since leaving zero offspring
implies that the genealogy goes extinct immediately. The second term
corresponds to the ancestor having one offspring, at Hamming distance
$i$ from the portal, that will {\em not} give rise to discovery of the
portal. That is, since this offspring itself forms the ancestor of a
new valley lineage, $\bar{p}_i(t-1)$ gives the probability that none
of {\em its} descendants discovers the portal within the next $t-1$
generations. The third term corresponds to the ancestor having two
offspring, one at distance $i$ from the portal and one at distance
$k$, neither of which give rise to the discovery of $\Omega$.  The
higher-order terms in Eq. (\ref{PrDiscoverPortal}) correspond to the
ancestor having $3$, $4$, and more offspring.

Recall that the mutation operator $\MuOp_{ij}$, as defined in Eq.
(\ref{MutationOperator}), gave the probability to go from Hamming
distance $j$ to distance $i$ from the {\em peak} under mutation.
$\MuOp_{ij}$ appears above in Eq. (\ref{PrDiscoverPortal}) with a
different, but equivalent, meaning: $\MuOp_{ij}$ there gives the
probability to go from a Hamming distance $j$ to a distance $i$
from the {\em portal} under mutation. This use of $\MuOp_{ij}$
appears repeatedly in the following.

Using Eq. (\ref{poiss_offspring}) we can sum the series in Eq.
(\ref{PrDiscoverPortal}), obtaining:
\begin{equation}
\bar{p}_j(t) =
  e^{\left( \left[ \bar{\bf p}(t-1) \cdot {\bf M} \right]_j -1\right)
  / \AvFit}
~,
\label{formal_bar_p_t}
\end{equation}
where
$\bar{p}(t) = ( \bar{p}_1(t), \bar{p}_2(t), \ldots, \bar{p}_L(t))$
and the vector notation denotes the sum
\begin{equation}
\left[ \bar{p}(t-1) \cdot {\bf M} \right]_j
  = \sum_{i=1}^L \bar{p}_i(t-1) \MuOp_{ij} ~.
\end{equation}

For $\AvFit \geq 1$, a valley genealogy eventually either
discovers $\Omega$ or goes extinct; see, for instance, Ref.
\cite{Harrisbook}. Letting $t \rightarrow \infty$ in Eq.
(\ref{formal_bar_p_t}), we obtain a set of nonlinear equations for
the asymptotic probabilities $\bar{p}_j$ that a genealogical bush,
whose founder started at Hamming distance $j$ from $\Omega$, goes
extinct before any of its descendants discovers the portal.
These are given by
\begin{equation}
\bar{p}_j =
  e^{\left( \left[ \bar{p} \cdot {\bf M} \right]_j -1\right)
  / \AvFit}
  ~ .
\label{formal_bar_p}
\end{equation}

Equations (\ref{formal_bar_p}) appear to be unsolvable in closed
analytical form. Their solutions may be numerically approximated in a
straightforward manner; for instance, by simply iterating
Eq. (\ref{formal_bar_p_t}). However, in the regime where $\mu$ is
small and $\AvFit$ is not too close to $1$, the
probabilities $\bar{p}_i$ are generally very close to $1$. In this
regime, one may expand Eq. (\ref{formal_bar_p}) to first order around
$\bar{p}_i =1$. To do this, we introduce the probabilities $\epsilon_i
= 1-\bar{p}_i$ that the portal {\em does} get discovered by the
lineage before it goes extinct. To first order in $\epsilon_i$ we
obtain from Eq. (\ref{formal_bar_p}) the equations given by
\begin{equation}
\epsilon_j = 1 - e^{-\MuOp_{0j} / \AvFit}
  \left( 1 -
  \frac{\left[ {\bf \epsilon} \cdot {\bf M} \right]_j}
  {\AvFit}
  \right) ~,
\end{equation}
where ${\bf \epsilon} = (\epsilon_1, \ldots, \epsilon_L)$.
These equations can be easily inverted, yielding:
\begin{equation}
\label{epsilons_inv_form}
\epsilon_j = \sum_{i=1}^L
  \left(1 - e^{-\MuOp_{0i} / \AvFit} \right)
  \left( {\bf I} - {\bf R} \right)^{-1}_{ij} ~,
\label{EpsilonFormula}
\end{equation}
where ${\bf I}$ is the identity matrix and the matrix ${\bf R}$ has
components
\begin{equation}
\label{r_def}
{\rm R}_{ij} = \frac{\MuOp_{ij}}{\AvFit}
  e^{ -\MuOp_{0j} / \AvFit} ~.
\label{EpsilonFormulaII}
\end{equation}
Note that the indices $i$ and $j$ in the matrices run from $1$
to $L$, corresponding to ancestors at Hamming distances between $1$
and $L$ from the portal. Note also that, by definition, $\epsilon_0 =1$.

To first order, Eqs. (\ref{epsilons_inv_form}) give the probabilities
$\epsilon_j$ that a valley lineage, founded by an ancestor at a
Hamming distance $j$ from $\Omega$, discovers the portal before the
lineage goes extinct. Now to calculate the barrier crossing time we
just have to determine the number of new valley lineages that are
founded per generation.

\subsection{Crossing the Fitness Barrier}

Every generation, $M$ individuals are selected in proportion to
their fitness. Each such selection may lead to the seeding of a new
lineage in the valley. The probability $P^{\rm not}$ that a selection
will {\em not} lead to the founding of a new valley lineage is given by:
\begin{equation}
\label{p_not}
P^{\rm not} = \frac{1-\PkPop}{\AvFit} + \frac{\sigma}{\AvFit}
  \PkPop (1-\mu)^L ~.
\end{equation}
The first term corresponds to selecting a valley individual, that by
definition is already on a lineage.
The second term corresponds to a peak individual being selected
and reproducing without mutation, leaving an offspring on the peak.

The probability $P^{\rm seed}_j$ that a new lineage will be seeded in
the valley and at a Hamming distance $j$ from $\Omega$ is given by 
\begin{equation}
P^{\rm seed}_j = \frac{\sigma}{\AvFit} \PkPop
  \left[ \MuOp_{jw} - (1-\mu)^L \delta_{jw} \right] ~,
\label{p_found_j}
\end{equation}
where $w$ is the Hamming distance between the peak and the portal. The
first factor, $\sigma P_{\Pi}/\AvFit$, gives the probability
that a peak individual will be selected. The term $\MuOp_{jw}$ is the
probability that under mutation this peak individual moves from Hamming
distance $w$ to distance $j$ from the portal. For $j \neq w$ this always
corresponds to a new lineage in the valley. For $j=w$, we must discount
for the probability that the peak individual did not undergo any
mutations at all. This is given by the term $(1-\mu)^L \delta_{jw}$.

Putting these together, we find the probability $\bar{P}^{\rm sel}$,
that a selection does {\em not} seed a lineage leading to the portal,
is given by
\begin{equation}
\bar{P}^{\rm sel} = P^{\rm not} + \sum_{j=1}^L (1-\epsilon_j) P^{\rm
  seed}_j.
\label{PrFound}
\end{equation}
Using Eqs. (\ref{formal_bar_p}), (\ref{p_not}), and (\ref{p_found_j})
and the identity $\AvFit = 1 + (\sigma-1)\PkPop$, Eq. (\ref{PrFound})
can be rewritten as
\begin{eqnarray}
\nonumber
\bar{P}^{\rm sel} = 1 & + & \frac{\sigma (\AvFit -1)}
  {(\sigma-1) \AvFit} \\
  & \times & \left[ (1-\mu)^L \epsilon_w + \AvFit \log(1-\epsilon_w) \right] ~.
\end{eqnarray}
Expanding the logarithm to first order in $\epsilon_w$, and using the
approximation in Eq. (\ref{av_fit_approx}) for $\AvFit$,
we obtain the simple expression
\begin{equation}
\bar{P}^{\rm sel} = 1 - \epsilon_w (\AvFit - 1).
\end{equation}

The probability that none of the $M$ selections from the current
generation seeds a lineage that discovers the portal is simply
$\left(\bar{P}^{\rm sel} \right)^M$. By our assumption of a
roughly constant quasispecies distribution, this probability is
constant. Thus, the expected number $\langle t \rangle$ of generations
until a lineage will be seeded that discovers the portal is given by
\begin{equation}
\label{cross_time}
\langle t \rangle = \frac{1}{1- \left(\bar{P}^{\rm sel}\right)^M}
\approx \frac{1}{M (\AvFit -1) \epsilon_w} ~.
\end{equation}
where $\epsilon_w$ is given by Eqs. (\ref{EpsilonFormula}) and
(\ref{EpsilonFormulaII}).

Equation (\ref{cross_time}) constitutes our theoretical prediction for
the average barrier crossing time $\langle t \rangle$ as a function of
the population size $M$, the fitness differential $\sigma$ between the
peak and the valley, the mutation rate $\mu$, the string length $L$,
and the width $w$ of the fitness barrier. To obtain it, we made
several approximations. We assumed that $\sigma$ was large enough and
$\mu$ small enough such that $\AvFit$ was substantially
larger than $1$. Under those assumptions, lineages in the valley are
short lived, the total number of individuals in the valley will be
small with respect to $M$, and the probabilities $\epsilon_i$ will be
small. This justifies our leading-order expansion for
$\langle t \rangle$.

\subsection{Additional Time in Valley Bushes}
\label{LengthValleyBushes}

Equation (\ref{cross_time}) gives the average number $\langle t \rangle$
of generations until a lineage is founded that discovers the portal.
The actual average time until the portal is discovered is somewhat
longer, since the lineage that finds the portal itself takes a certain
average number of generations to discover the portal. Specifically,
there is an additional average time, that we denote by
$\langle d t \rangle$, between the {\em founding} of the first lineage
that discovers the portal and the actual discovery of $\Omega$.

We can directly approximate this
correction term $\langle dt \rangle$ when the $\epsilon_i$ are
small. As we will see below, generally $\langle dt \rangle \ll \langle
t \rangle$ in the parameter regime where our approximations are
valid. This makes the effect of including the correction term $\langle
dt \rangle$ rather small in these parameter regimes. However, as we
approach the parameter regime where the $\epsilon_i$ become large, the
average number of generations $\langle t \rangle$ until the founding
of the lineage that discovers the portal becomes comparable to the
average number of generations $\langle d t \rangle$ that it takes this
lineage to actually discover the portal. In this (limited) parameter
regime, including the correction term $\langle dt \rangle$ leads to a
significant improvement of our theoretical predictions.

Paralleling the development of the Eq. (\ref{formal_bar_p}), we start
by expanding Eq. (\ref{formal_bar_p_t}) to first order in
$\epsilon_j(t)$; the probability that the lineage starting at distance
$j$ has discovered the portal by time $t$. We find that
\begin{equation}
\label{eps_t}
\epsilon_j(t) = 1 -
  e^{-\MuOp_{0j} / \AvFit}
  \left( 1- \frac{ \left[ {\bf \epsilon}(t-1) \cdot {\bf M} \right]_j}
  {\AvFit} \right) ~.
\end{equation}
The expected additional time $\langle dt_j \rangle$, given that the
lineage started at a Hamming distance $j$ from $\Omega$ and
{\em conditioned} on the lineage discovering the portal, is formally
given by
\begin{eqnarray}
\nonumber
\langle dt_j \rangle & = & \sum_{t=1}^{\infty} t \;
  \frac{\epsilon_j(t)-\epsilon_j(t-1)}{\epsilon_j} \\
  & = & \sum_{t=0}^{\infty} \frac{\epsilon_j-\epsilon_j(t)}{\epsilon_j}
  ~,
\end{eqnarray}
where the asymptotic $\epsilon_j$ is given by Eqs. (\ref{EpsilonFormula})
and (\ref{EpsilonFormulaII}). Using Eq. (\ref{eps_t}) and the boundary
conditions $\epsilon_j(0) = 0$, the above sum gives:
\begin{equation}
\label{dt_j_expression}
\langle dt_j \rangle = \frac{1}{\epsilon_j}
  \sum_{i=1}^L \left( 1- e^{-\MuOp_{0i} / \AvFit} \right)
  \left( {\bf I} - {\bf R} \right)^{-2}_{ij},
\end{equation}
where the matrix ${\bf R}$ is again defined by Eq. (\ref{r_def}).


In order to obtain $\langle dt \rangle$ we have to {\em weigh} each of
the times $\langle dt_j \rangle$ with a factor $c_j$ corresponding
to the relative proportion of times that a lineage starting at Hamming
distance $j$ discovers the portal. That is, averaged over an ensemble
of runs, $c_j$ is the proportion of times that the portal was discovered
by a lineage that started at Hamming distance $j$. The weight $c_j$
should be proportional to both the probability $\epsilon_j$ and the
rate of creating of lineages at Hamming distance $j$ from the portal.
We have that
\begin{eqnarray}
\nonumber
c_j = \frac{\epsilon_j \left(\MuOp_{jw} - (1-\mu)^L \delta_{jw}
  \right)}{\sum_k \epsilon_k \left(\MuOp_{kw} - (1-\mu)^L
  \delta_{kw}\right)} ~, \\
  j = 0,1, \ldots, L ~,
\label{c_j_expression}
\end{eqnarray}
where the factors in parentheses are similar to that found in Eq.
(\ref{p_found_j}).
It should be noted that here the indices run from $0$ to $L$ and
not from $1$ to $L$, since the portal may also be discovered by a jump
mutation directly from the peak.

Combining Eqs. (\ref{dt_j_expression}) and (\ref{c_j_expression}) and
using Eq. (\ref{epsilons_inv_form}), we find that the average length
of the valley bush that discovers the portal is
\begin{equation}
\label{dt_expression}
\langle dt \rangle = \frac{ \left[{\bf \epsilon} \cdot \left( {\bf I} - {\bf
      R}\right) \cdot \left( {\bf M} - {\bf I} (1-\mu)^L
      \right)\right]_w}{\left[ {\bf \epsilon} \cdot \left( {\bf M} - {\bf I}
      (1-\mu)^L \right)\right]_w + \MuOp_{0w}} ~,
\end{equation}
where, again, $\bf \epsilon$ is given by its components in Eqs.
(\ref{EpsilonFormula}) and (\ref{EpsilonFormulaII}). The indices in the
vector notation now run from $1$ to $L$.

Adding the correction term $\langle dt \rangle$ to $\langle t \rangle$
as given by Eq. (\ref{cross_time}) improves our theoretical predictions
especially in the regime where the $\epsilon_i$ become large. However,
we still expect the approximations leading to the above equations for
$\langle t \rangle$ and $\langle dt \rangle$ to break down when
$\AvFit \rightarrow 1^+$.

\subsection{Theory versus Simulation}

\begin{figure}[htbp]
\centerline{\epsfig{file=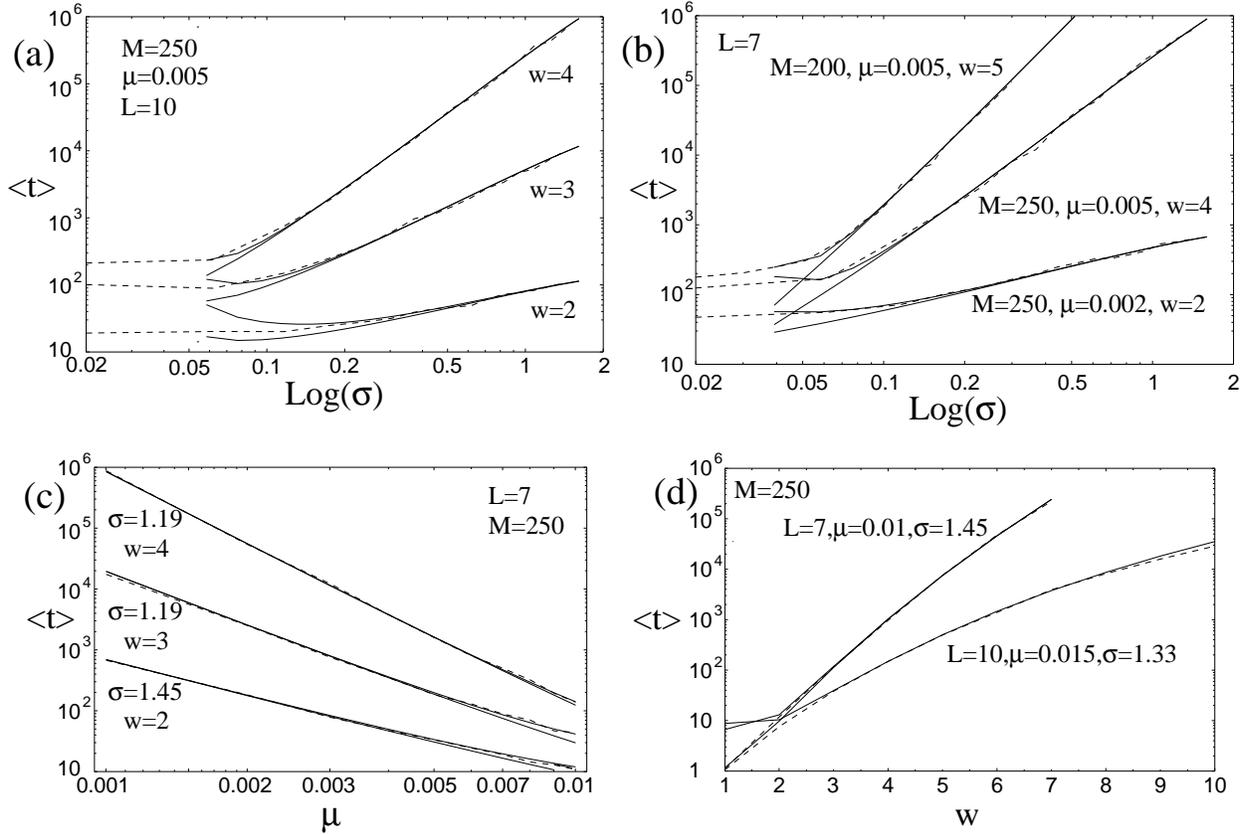,width=6.5in}}
\caption{Barrier-crossing times $\BCTime$ as a function of
  barrier height $\sigma$, mutation rate $\mu$, and barrier width $w$,
  for a variety of parameter settings. The simulation results are
  plotted using dashed lines. Each point on each dashed line is an
  estimate of $\langle t \rangle$ averaged over $250$ simulation runs.
  The theoretical predictions are shown as pairs of solid lines: The
  lower of each pair gives the theoretical predictions of Eq.
  (\ref{cross_time}), while the higher has the additional correction
  term of Eq. (\ref{dt_expression}) added. Note that, except for the
  horizontal axis in Fig. (d), all axes use a logarithmic scale.
  General parameter settings are indicated at the top of each plot,
  while parameters specific to the different runs are indicated next
  to the their lines.}
\label{theo+exp_fig}
\end{figure}

We simulated an evolving population using a fitness function
consisting of a single barrier, as described in
Secs. \ref{evolutionary_dynamics} and \ref{crossing_single_barrier},
for a wide range of parameter settings to quantitatively test our
theoretical predictions. Results for several parameter regimes are
shown in Fig. \ref{theo+exp_fig}, where the simulation results are
plotted using dashed lines and the theoretical predictions are plotted
with solid lines. Each data point on the dashed lines was obtained by
averaging over $250$ runs with equal parameter settings. The
theoretical predictions are shown as pairs of solid lines, where the
lower solid line in each pair shows the predictions from
Eq. (\ref{cross_time}) and the upper solid line shows
Eq. (\ref{cross_time}) plus the correction term of Eq.
(\ref{dt_expression}). Note that for most parameter ranges the
difference between the two solid lines is so small as to be
undetectable.

Figures \ref{theo+exp_fig}(a) and \ref{theo+exp_fig}(b) show the
average barrier crossing time $\langle t \rangle$ as a function of the
logarithm $\log (\sigma)$ of the barrier height. Additionally, both
$\langle t \rangle$ and $\log (\sigma)$ are plotted using a
logarithmic scale. The shapes of the curves correspond to the
dependencies of $\log \langle t \rangle$ on $\log ( \log \sigma
)$. Portions of curves that are straight lines thus indicate a scaling
of the form $\langle t \rangle \propto (\log \sigma)^s$, with $s$ the
slope of the straight portion. Note that $\langle t \rangle$ ranges
over $5$ orders of magnitude, from $10$ to $10^6$, in both
Figs. \ref{theo+exp_fig}(a) and \ref{theo+exp_fig}(b). We see that the
theory accurately predicts the simulation results for barrier heights
that are not too small.

\end{multicols}

\begin{multicols}{2}

In Fig. \ref{theo+exp_fig}(a) the theory starts deviating from the
experimental data around $\log(\sigma) \approx 0.06$ for the upper two
curves and around $\log(\sigma) \approx 0.15$ for the lowest. These values
of $\sigma$ correspond to selective advantages $\sigma-1$ of the
peak of a little over $6$ and $16$ percent, respectively. Notice that
the upper two experimental curves are almost horizontal for small
values of $\sigma$ up to $\log(\sigma) \approx 0.06$, after which they trend
upwards becoming almost linear. As we will show below, it turns out
that the location of this crossover is found at the {\em
finite-population error threshold} that separates the entropy-barrier
regime from the fitness-barrier regime. That is, for the parameters
$M=250$, $\mu =0.005$, and $L=10$, the critical value $\sigma_c$ below
which the population dynamics acts effectively as if there were no
fitness peak at all occurs around $\log(\sigma_c) \approx 0.06$. The
same phenomenon is observed in the two upper curves of Fig.
\ref{theo+exp_fig}(b): the crossover occurs around
$\log(\sigma_c) \approx 0.05$. Note that the correction terms $\langle
dt \rangle$ extend the parameter region over which the theory provides
accurate predictions approximately up to the finite-population error
threshold.

Above the error threshold, for values of $\sigma$ in the fitness-barrier
regime, the curves appear nearly linear. This indicates that the barrier
crossing times scale with powers of the logarithm of the barrier height
$\sigma$: $\langle t \rangle \propto (\log \sigma)^s$, where $s$ is the
line's slope. Thus, the barrier crossing time increases relatively slowly
as a function of the barrier height. One further observes that the
barrier crossing times are not only longer for wider barriers (larger
values of $w$), but that the slopes of the curves are larger as well.
That is, for large widths $w$ the barrier crossing time increases faster
as a function of $\sigma$ than for low values of $w$.

Figure \ref{theo+exp_fig}(c) shows the barrier crossing time $\langle t
\rangle$ as a function of the mutation rate $\mu$, for three different
values of the barrier width $w$ and two different values of the
barrier height $\sigma$. The population size is $M=250$ and the
genotypes length $L=7$ for all three curves. On the logarithmic scales,
the curves again look approximately linear, indicating that the
barrier crossing time scales as a power law in the mutation rate
$\mu$: $\langle t \rangle \propto \mu^s$, where $s$ is the slope. We
again see that for wider barriers, the waiting times are both larger
and vary more rapidly with $\mu$. The theory predicts the simulation
results quite accurately over the entire range. Only for large mutation
rates ($\mu \approx 0.01$) do the theoretical predictions with and without
the correction term of Eq. (\ref{dt_expression}) differ significantly.
In this regime the theoretical and experimental values start to differ
slightly as well, although the predictions are still accurate. It is
notable in the two lower curve families, with barrier widths $w=2$ and $w=3$,
that the correction term $\langle dt \rangle$ improves the theoretical
predictions for high mutation rates.

Finally, Fig. \ref{theo+exp_fig}(d) shows the barrier crossing time
$\BCTime$ as a function of the barrier width $w$. Only the barrier
crossing time $\BCTime$ is shown on a logarithmic scale, so that any
linear dependence indicates an exponential scaling: $\BCTime \propto
10^{sw}$, where $s$ is the slope. Again, the theory accurately
predicts the barrier crossing time. The fact that the curves are not
linear and bend downwards shows that $\BCTime$ grows more slowly than
exponential with barrier width; although it still increases rapidly as
a function of $w$. In fact, we chose large values of the mutation rate
$\mu$ in these plots ($\mu = 0.01$ and $\mu = 0.015$) to ensure that
the barrier crossing time is still in a reasonably bounded range up to
large barrier widths. For smaller mutation rates, the barrier crossing
times become so large as to make it impossible to perform an adequate
number of simulation runs. For the case $w=1$, the correction term
$\langle dt \rangle$ leads to an overestimation of $\BCTime$. Note, however,
that for $w=1$ there is effectively no fitness barrier; the portal is a
mutant neighbor of the peak genotype and so valley bushes are
essentially nonexistent.

In summary, Figs. \ref{theo+exp_fig}(a)-(d) show that the theoretical
predictions of Eq. (\ref{cross_time}), possibly including the correction
term of Eq. (\ref{dt_expression}), accurately predict the average
barrier crossing times estimated over a wide range of parameters from
simulations of an evolving population. The theory breaks down, as expected,
when the barrier height $\sigma$ becomes small ($\sigma \approx 1$)
and this is illustrated on the left-hand sides of Figs.
\ref{theo+exp_fig}(a) and \ref{theo+exp_fig}(b). In this low-$\sigma$
regime, which sets in suddenly as a function of $\sigma$, $\BCTime$
becomes almost independent of $\sigma$. Roughly speaking, the selection
pressure is too small to keep the population concentrated around the peak,
and the population randomly diffuses through the valley until it discovers
the portal. In this regime, the barrier is in effect not a fitness barrier,
but an entropy barrier.

In Sec. \ref{err_thres_and_neut_reg} we will analyze
the location of the finite-population error threshold that separates
the fitness and entropy barrier regimes and discuss the entropy-barrier
crossing population dynamics. In the next subsection, though, we first
discuss the scaling of the fitness-barrier crossing time $\BCTime$ with
the different parameters $\sigma$, $w$, $\mu$, and $M$.

\subsection{Scaling of the Barrier Crossing Time}
\label{scaling_bar_cross_time}

In Fig. \ref{theo+exp_fig} we saw, by varying one parameter at a time,
that the barrier crossing time scaled as a power law in the logarithm
of the barrier height $\log(\sigma)$, as a power law in mutation rate
$\mu$, and somewhat slower than exponential in the barrier width $w$.
Analytically extracting these scalings from Eq. (\ref{cross_time}) is
quite challenging and incomplete at this time. Empirically, though, we
found that the barrier crossing time can be fit quite accurately, in
the regime where $\sigma$ is not too small (above the error
threshold), to a scaling function with the following form
\begin{eqnarray}
\label{empirical_scaling}
\langle t \rangle \propto \frac{1}{w! M \mu} &&
  \left( \frac{\log(\sigma)}{\mu}\right)^{w-1} \\ \nonumber
  & \times & \left[ \log(\sigma) \right]^{-\gamma -\delta \log(\mu)} ~,
\end{eqnarray}
where $\gamma$ and $\delta$ are (constant) scaling exponents. For {\em
both} the genotype lengths ($L=7$ and $L=10$) for which we have
detailed data, we found that $\gamma \approx 0.75$ and
$\delta \approx 0.1$.

This empirical scaling law confirms that, in fact, the
barrier crossing time $\BCTime$ scales as a power law in both
$\log(\sigma)$ and $\mu$.  We see, in particular, that the dependence
on the mutation rate $\mu$ scales roughly inversely with $\mu^w$ and
the dependence on $\log(\sigma)$ scales roughly as $\left[
\log(\sigma)\right]^{w-2}$. Furthermore, we see that $\BCTime$ scales
as $e^{c w}/w!$, with $c$ a constant, when only the barrier width $w$
is varied. The scaling with $w$ is thus by far the {\em most rapid}
and therefore dominant scaling. That is, widening the barrier
increases the waiting time $\BCTime$ much more than increasing the
height of the barrier or decreasing the mutation rate.

These empirically observed scaling behaviors can be elucidated using a
simple analytical argument. To this end we employ several
simplifications. First, we assume that the major contributions to the
probability of barrier crossing come from terms with the {\em minimal}
number of mutations. That is, for barriers of width $w$, at least $w$
mutations must occur in a peak individual in order to discover the
portal. Thus, we assume that contributions from ``paths'' between peak
and portal that involve more than $w$ mutations are negligible.  This
for instance implies that we neglect the contributions from lineages
founded at Hamming distances $w$ through $L$ from $\Omega$.
Furthermore, we assume that valley lineages are unlikely to be founded
more than $1$ mutation away from the peak. Putting these together, the
dominant contribution to the barrier crossing probability comes from
lineages that are founded at a Hamming distance $w-1$ from the
portal. Note that if we set $P_{\Pi} = 1$ for simplicity, each
generation approximately
\begin{equation}
\frac{w \mu M}{1-\mu}
\end{equation}
such lineages are founded.

We will now estimate the probability that a lineage, starting at
Hamming distance $w-1$ from the portal, discovers the portal exactly
$t$ generations after its founding.  We approximate the valley
genealogies by assuming that each valley individual can only have zero
or one offspring each generation. This implies that a valley lineage
consists of a single line of individuals; i.e., lineages do not
branch. The probability that such a lineage persists for at least $t$
time steps is $1/\AvFit^t$. At $t=0$, the lineage has $w-1$ bits set
incorrectly, and $L-w+1$ bits set correctly. In order for the lineage
to discover the portal {\em exactly} at time $t$, it will have to
mutate its bits such that, at time $t$ and for the first time, the
$w-1$ ``incorrect'' bits will all have been flipped to the correct
state and all the $L-w+1$ correct bits are left undisturbed. Thus,
between time $0$ and $t$, the $w-1$ incorrect bits have to be mutated
exactly once, while the correct bits have been undisturbed. Since we
are calculating the probability for the portal to be discovered
exactly at time $t$, one of the $w-1$ bits has to flip at time $t$,
while the other $w-2$ might flip at any prior time. This gives $(w-1)
\: t^{w-2}$ possibilities for contributing flips. All other bits have
to remain unflipped for all time steps.

Thus, the probability $P^{{\rm find}}_t$ that a lineage finds the
portal exactly at time $t$ is approximately given by:
\begin{eqnarray}
\nonumber
P^{{\rm find}}_t & = & (w-1) \: t^{w-2} \\
  & \times & \mu^{w-1} (1-\mu)^{L t - w+1}
  \left( \frac{1}{\AvFit}\right)^t ~,
\label{PrFind}
\end{eqnarray}
where the last factor gives the probability that the lineage survives
until time $t$. Using Eq. (\ref{av_fit_approx}) and summing Eq.
(\ref{PrFind}) over $t$ we find:
\begin{eqnarray}
\nonumber
P^{{\rm find}} & = & \sum_{t=0}^\infty P^{\rm find}_t \\ \nonumber
  & = & (w-1) \left(\frac{\mu}{1-\mu}\right)^{w-1} \sum_{t=0}^{\infty}
  t^{w-2} \left( \frac{1}{\sigma} \right)^t \\
  & = & (w-1) \left(\frac{\mu}{1-\mu}\right)^{w-1}
  {\rm Li}_{2-w}\left(\frac{1}{\sigma}\right) ~,
\end{eqnarray}
where ${\rm Li}_n(x)$ is the poly-logarithm function: essentially
defined by the sum in the second line above. It is more insightful to
approximate the sum with an integral. We then obtain
\begin{eqnarray}
\nonumber
P^{{\rm find}} & = & \left(\frac{\mu}{1-\mu}\right)^{w-1}
  \int_{0}^{\infty} (w-1) \: t^{w-2} e^{-\log(\sigma) t} dt \\
  & = & (w-1)! \left(\frac{\mu}{\log(\sigma) (1-\mu)}\right)^{w-1} ~.
\end{eqnarray}

Recall that the rate at which lineages at Hamming distance $w-1$ are
being created is $w \mu M/(1-\mu)$. Using this and noting that the
barrier crossing time is inversely proportional to $P^{\rm find}$,
we obtain a scaling of the form
\begin{equation}
\langle t \rangle \propto \frac{1}{w! M \mu}
  \left( \frac{\log(\sigma)}{\mu}\right)^{w-1} ~,
\label{analytic_scaling}
\end{equation}
where we have neglected the factor $(1-\mu)^w \approx 1$. The scaling
relation of Eq. (\ref{analytic_scaling}) recovers most of the
empirically determined scaling behavior in Eq.
(\ref{empirical_scaling}).

The dominant scaling with $\mu$ and $\log(\sigma)$ can be understood
as follows. The average time that a lineage spends in the valley
before going extinct is roughly $1/\log(\sigma)$. Thus,
$\mu/\log(\sigma)$ gives the average number of mutations that a
lineage in the valley undergoes before it goes extinct. Since this
number is generally much smaller than $1$, it can be interpreted as
the probability of having a single mutation in a valley lineage. The
probability of having $w-1$ mutations is then of course
$(\mu/\log(\sigma))^{w-1}$. There is an additional factor $1/\mu$ from
the rate at which valley lineages are being created at Hamming
distance $w-1$. Ref. \cite{Weis91a} also argues, along somewhat
different lines, that the barrier crossing time should have a
power-law dependence on mutation rate: $\langle t \rangle
\propto \mu^{-w}$.

The correction factors---those with scaling
exponents $\gamma$ and $\delta$ in Eq. (\ref{empirical_scaling})---
probably arise from the fact that lineages are not simple unbranching
lines of descendants, as we have assumed, but are more complicated
tree-like genealogies.

The factor $w!$ in Eq. (\ref{analytic_scaling}) counts the number of
distinct paths of minimal length between the peak and the portal.
Curiously, it appears from the scaling formulas that when $w$ gets very
large, the barrier crossing time starts to {\em decrease} again.
Applying Stirling's approximation to the factorial function in
Eq. (\ref{analytic_scaling}) indicates that $\BCTime$ has a maximum
around $w \mu = \log(\sigma)$. Although this may initially seem
strange, it does make sense, since as we will now argue, fitness
barriers for which $w \mu > \log(\sigma)$ do not exist.

If there are $w!$ independent paths between peak and portal, this
implies that there are $w$ independent directions from the peak into
the valley. In other words, at least $w$ bits of the peak genotype can
undergo deleterious mutations. As we will see in
Sec. \ref{err_thres_and_neut_reg} below, the error threshold at which
peak individuals becomes unstable in the population occurs near
\begin{equation}
\sigma (1-\mu)^{L} = 1 ~.
\end{equation}
To first order in $\mu$, this is equivalent to $\log(\sigma) = L \mu$.
That is, if $L \mu  > \log(\sigma)$, peak individuals will be lost from
the population, and the population will start diffusing randomly through
genotype space. Obviously, the genotype length has to be longer than
the barrier width $L > w$. Therefore, $w \mu > \log(\sigma)$ implies
that the genotype length $L$ is so large that it is impossible to
stabilize the peak individuals. In other words, fitness barriers with
$w \mu > \log(\sigma)$ simply do not exist.

Finally, it should be noted that as $\log(\sigma) \rightarrow \infty$,
the barrier crossing time goes to a finite asymptote and {\em not} to
infinity. Since valley lineages have probability zero to reproduce in
this limit, the asymptotic barrier crossing time is given by the (finite)
waiting time for a ``long jump'' in which a peak individual undergoes
$w$ mutations at once.

The main consequence of the scaling relations just derived, is that if
the population is located on a fitness ``plateau'' in genotype space,
surrounded by different valleys on all sides, then it will most likely
escape from the plateau via the valley with the smallest {\em width}
and {\em not} via the valley path with the smallest depth. One
concludes that high barriers can be passed relatively easily, as long
as they are narrow; while wide barriers take a very long time to
cross, even if they are shallow.

We should emphasize that this situation is very different from the
scaling of barrier crossing times generally encountered in physics or,
for that matter, in evolutionary models that literally interpret the
``landscape'' metaphor as leading to stochastic gradient dynamics on a
fitness ``potential''. In these settings, the system's state space has
an energy function defined on it that acts as a potential field. In
the absence of any noise, the system is assumed to follow the gradient
(downward) of the energy ``landscape''. In the presence of noise, the
system can deviate from its gradient path, but movement against the
gradient is unlikely in proportion to its deviation from the local
gradient. The barrier crossing times then depend mainly on the barrier
{\em height}, and they scale exponentially with this barrier height
\cite{Gardiner85}.

For example, imagine an energy barrier that consists of a steep slope
upwards, followed by a long plateau and then a steep slope leading
downward on the other side. The initial steep ascent from the valley
onto the plateau is very unlikely since it involves moving against a
steep gradient. However, after this unlikely step has been
established, the system can cross the long plateau to the other side
relatively easily, since it does not involve moving against an energy
gradient. Thus, the width of the energy barrier is almost immaterial,
while the barrier height is the defining impediment, since it
determines the extent to which movement against the gradient must
occur.

The situation is entirely different for fitness ``landscapes'' in
which an evolving population moves. For an evolving population, making
a large jump in fitness is not unlikely at all. One mutation in one
individual can do the trick. However, since some individuals remain at
the peak, the individuals in the valley are continuously in
competition with these higher-fitness peak individuals. An absolute
fitness scale is set by these peak individuals. It is therefore
survival at low fitness---compared to the most-fit individuals in the
population---for an extended period of time that is unlikely. And this
is why the time it takes to move across the plateau is the key
parameter---which, of course, is controlled by the barrier width.

The preceding discussion should make it clear, once again, that this
analogy---between a population evolving over a fitness ``landscape''
and a physical system moving over its energy ``landscape'' in state
space---is problematic: at best it may lead one to the wrong intuitions;
at worst the basic physical results simply do not describe
evolutionary behavior.

\section{The Entropy Barrier Regime}
\label{err_thres_and_neut_reg}

Figures \ref{theo+exp_fig}(a) and \ref{theo+exp_fig}(b) showed that
below a critical barrier height $\sigma_c$, where the dashed lines
began to run horizontally, the barrier crossing time became
effectively independent of $\sigma$. We also saw that the theory
breaks down for $\sigma < \sigma_c$. In this regime, all peak
individuals are quickly lost and the population diffuses through
the valley until the portal is discovered. The theoretical
calculations, in contrast, assumed the population was located at the
peak and that short lineages were continuously spawned in the
valley. It is no surprise then that the predictions break down in this
regime. Since the population dynamics is dominated by diffusing
through the valley's fitness-neutral volume, we refer to this as the
{\em entropy-barrier regime}. Before discussing the barrier crossing
times in this entropic regime, we first estimate the error threshold's
location $\sigma_c$ as a function of $\mu$, $M$, and $L$.

\subsection{Error Thresholds}
\label{ErrorThresholds}

As a first, population-size independent approximation one might guess
that the error threshold occurs when the average number of
peak-offspring produced by a single peak individual in a population of
valley individuals is $1$. From Eq. (\ref{av_fit_approx}) this leads to
an estimated critical barrier height $\sigma_c$ of
\begin{equation}
\label{crit_sigma_infpop}
\sigma_c = (1-\mu)^{-L} ~.
\end{equation}
This equation is the standard error-threshold result in molecular
quasispecies theory \cite{Eigen71,Eigen&McCaskill&Schuster89}. For the
parameters $L=10$ and $\mu =0.005$ of Fig. \ref{theo+exp_fig}(a), this
leads to $\log(\sigma_c) \approx 0.05$, or $\sigma_c \approx 1.05$. As
seen from the figure, though, the entropic regime extends to somewhat
higher peak fitness; as far as $\log(\sigma) \approx 0.06$. This
deviation is due to finite-population sampling effects and to
neglecting back mutations, which become important as the error
threshold is approached.

For finite populations, the error threshold can be defined most
naturally as those parameter values for which the mean proportion
$\PkPop$ of peak individuals equals the variance, due to finite
population fluctuations, in $\PkPop$. That is, the criterion for
reaching the error threshold is 
\begin{equation}
\label{error_thres_def}
(\PkPop)^2 = {\rm Var} (\PkPop) ~.
\end{equation}
The intuition behind this definition is as follows: Since the
proportion of peak individuals fluctuates, eventually a large
fluctuation will occur that leads to the loss of all peak
individuals. As was shown in Ref. \cite{Nimw97b}, however, the
waiting time for such a destabilization to occur increases
exponentially with the ratio $(\PkPop)^2/{\rm Var}(\PkPop)$. Only when
$(\PkPop)^2 < {\rm Var}(\PkPop)$ do such destabilizations occur
relatively frequently. For $(\PkPop)^2 > {\rm Var}(\PkPop)$ the
fluctuations are small enough so that the proportion of peak
individuals typically does not vanish. Therefore, it is natural
to use Eq. (\ref{error_thres_def}) to delineate the regimes with
``unstable'' and ``stable'' peak populations and so to distinguish
between the fitness-barrier and entropy-barrier regimes in the
population dynamics.

Finite-population error thresholds may also be defined in alternative
ways; cf. Ref. \cite{Nowak&Schuster}. Typically, one finds that,
although the conceptual motivations differ, the quantitative parameter
values for which the different error thresholds occur are quite similar.

The variance ${\rm Var} (\PkPop)$ can be most easily calculated using
diffusion-equation methods. For an introduction to these techniques in
the context of mathematical population genetics, see for instance
Ref. \cite{Kimura64}. To begin, we assume that, due to sampling
fluctuations, at some particular time $t$ the actual proportion of
peak individuals is not $\PkPop$ but instead is $P(t) = \PkPop + x(t)$.
That is, the proportion $P(t)$ of individuals on the peak deviates
$x(t)$ from its equilibrium value $\PkPop$. We focus on the dynamics of
the deviation $x(t)$. At the next generation, the expected deviation
$\langle x(t+1)\rangle$ is
\begin{equation}
\langle x(t+1) \rangle = \frac{x(t)}{\AvFit} ~.
\end{equation}
Thus, the {\em expected} change $\langle \delta x \rangle$ in the
deviation is given by
\begin{equation}
\langle \delta x \rangle = \frac{1-\AvFit}{\AvFit}
 \: x \equiv -\gamma \: x ~,
\label{ChangeMean}
\end{equation}
where we have defined $\gamma$ by the last equality. $\gamma$ measures
the average rate at which fluctuations around the quasispecies
equilibrium distribution are damped. The second moment $\langle
\left(\delta x\right)^2 \rangle$ of the change $\delta x$ is
approximately given by the variance of the binomial-sampling
distribution. One finds that
\begin{equation}
\langle \left(\delta x\right)^2 \rangle = \frac{1}{M} \left(\PkPop +
\frac{x}{\AvFit} \right) \left(1-\PkPop - \frac{x}{\AvFit} \right).
\label{ChangeVariance}
\end{equation}

A Fokker-Planck diffusion equation approximation determines the temporal
evolution of distribution ${\rm Pr}(x,t)$ of $x(t)$ via
\begin{eqnarray}
\nonumber
\frac{\partial}{\partial t} {\rm Pr}(x,t) =
  & - & \frac{\partial}{\partial x} 
  \langle \delta x \rangle {\rm Pr}(x,t) \\
  & + & \frac{1}{2} \frac{\partial^2}{\partial x^2}
  \langle (\delta x)^2 \rangle {\rm Pr}(x,t) ~,
\end{eqnarray}
where $\langle \delta x \rangle$ from Eq. (\ref{ChangeMean}) gives the drift
term and $\langle ( \delta x)^2 \rangle$ from Eq. (\ref{ChangeVariance}) 
the diffusion term. Solving for the limit distribution ${\rm Pr}(x)$ for $x$
yields
\begin{eqnarray}
\nonumber
{\rm Pr}(x) & = & C \left( \PkPop +\frac{x}{\AvFit}\right)^{2 M \AvFit
 (\AvFit -1) \PkPop} \\
  & \times & \left( 1 - \PkPop -\frac{x}{\AvFit}\right)^{2 M \AvFit
  (\AvFit -1) (1-\PkPop)} ~.
\end{eqnarray}
Here $C$ is a normalization constant that ensures ${\rm Pr}(x)$ is
normalized on the interval $x \in [-\PkPop,1-\PkPop]$. If we expand
the fluctuations to second-order around $x = 0$, the distribution
becomes a Gaussian given by
\begin{equation}
{\rm Pr}(x) = \tilde{C} e^{-\frac{M \gamma}{\PkPop(1-\PkPop)} x^2},
\end{equation}
where $\tilde{C}$ is again a normalization constant. From this
distribution of fluctuations one directly reads off the variance
${\rm Var}(\PkPop)$, finding that
\begin{eqnarray}
\nonumber
{\rm Var}(\PkPop) & = & \frac{\PkPop(1-\PkPop)}{2 M \gamma} \\
  & = & \frac{\AvFit \left( \sigma - \AvFit \right)}{2 M (\sigma -1)^2}
  ~,
\label{VarPkPop}
\end{eqnarray}
where we used Eqs. (\ref{PkIndividuals}) and (\ref{ChangeMean}) to
arrive at the last line.

As noted before, we define the finite-population error threshold by
those parameter values for which $(\PkPop)^2 = {\rm Var}(\PkPop)$.
Using Eq. (\ref{VarPkPop}) leads to the error-threshold parameter
constraints given by
\begin{equation}
\frac{2 M (\AvFit -1)^2}{\AvFit (\sigma - \AvFit)} = 1.
\label{fin_pop_erthres}
\end{equation}
If we substitute the parameter values $\mu = 0.005$, $M=250$, and
$L=10$ of Fig. \ref{theo+exp_fig}(a) and use Eq. (\ref{av_fit_approx})
for $\AvFit$, we find the error threshold at
$\log(\sigma_c) \approx 0.059$. This agrees quite well with the
location at which the experimental curves start bending upwards with
increasing peak height. 

\subsection{The ``Landscape'' Regime}

As we have pointed out previously, the scaling relations derived in
Sec. \ref{scaling_bar_cross_time} contrast strongly with those
based on ``landscape'' models in which the population {\em as a
whole} diffuses through the fitness landscape
\cite{Lande85,Newman&Cohen&Kipnis85}. For those models, the barrier
crossing time scales exponentially with population size and barrier
height. It turns out that this scaling behavior---appropriate
to the ``landscape'' regime---can be reconciled with the scaling
formulas derived in Sec. \ref{scaling_bar_cross_time} by closer
inspection of Eq. (\ref{fin_pop_erthres}).

As noted above, the average {\em destabilization time} for a
fluctuation to occur that makes all peak individuals disappear from
the population scales exponentially in the ratio ${\rm
Var}(P_{\Pi})/(P_{\Pi})^2$ given by
Eq. (\ref{fin_pop_erthres}). Thus, Eq. (\ref{fin_pop_erthres}) shows
that this destabilization time increases exponentially with population 
size. For cases where $\langle f \rangle \gg 1$ and for reasonable
population sizes, the destabilization time is so large that the barrier
crossing time is determined by how long it takes a rare mutant to
cross the fitness valley.

Close to the finite-population error-threshold
($\langle f \rangle \approx 1$), however, it might be the case that
the time to create such a rare sequence of mutants is long in
comparison to the destabilization time. In this situation, the barrier
crossing time is essentially given by the destabilization time: As
soon as all peak individuals are lost, the population diffuses through
the valley and quickly discovers the portal. Thus, in the very
restricted ``landscape'' parameter regime just around the
error-threshold, the barrier crossing time is determined by the
destabilization time and {\em does} scale exponentially with population
size and barrier height.

Beyond the error threshold---that is, for smaller populations, larger
mutation rates, smaller barrier heights, or longer genotypes---the
peak readily becomes unoccupied. In this regime, the barrier crossing
time becomes almost independent of barrier height $\sigma$. The
barrier to be crossed is then no longer a fitness barrier. Instead, it
has become an entropy barrier. The population must search through
almost all of the valley until the portal is discovered. Thus, only for
parameters near the boundary between the fitness and entropic
regime does the barrier crossing time scale in accordance with the
``landscape'' models. 

\subsection{Time Scales in the Entropic Regime}

The population dynamics in the entropic regime beyond the error
threshold is modeled most directly by considering an entirely flat
(constant) fitness function; in particular, one in which all genotypes
have fitness $1$ and containing a single portal $\Omega$. The
population starts out concentrated on a genotype at Hamming distance
$w$ from $\Omega$ and evolves under selection and mutation until the
portal genotype is discovered for the first time.  Denote this average
entropy-barrier crossing time by $\tau$.

The calculation of the entropy-barrier crossing time appears less
analytically tractable than the calculation of the fitness-barrier
crossing time. The main difficulty arises from the sampling of
individuals at each generation, combined with the global constraint
of a fixed population size. Due to this sampling dynamics, subtle
genetic correlations emerge between the individuals. Although some
of the aspects of the correlation statistics have been derived
analytically \cite{Derrida&Peliti}, the entropy-barrier crossing time
$\tau$ depends in a complicated, and not yet well understood, way on
these correlations. We will discuss the difficulties with calculating
entropic barrier crossing time by deriving several simple
approximations and discussing why they fail to provide accurate
quantitative predictions.

First, one can approximate the neutral evolution just defined by
assuming that each individual in the population has exactly one
offspring. In this case, the population effectively consists of $M$
independent {\em random walkers} that diffuse through genotype
space. Since each individual has only one offspring one can identify
its genealogy with a single evolving genotype that mutates each bit
with probability $\mu$ at each generation. Since $\mu L \ll 1$ in
general, this genotype effectively performs a random walk in the
hypercube, where random walk steps are made at a rate of one step per
$1/(L \mu)$ generations on average.

The average time $\tau_1$ a {\em single} random walker takes to
discover $\Omega$ is given by:
\begin{equation}
\tau_1 = \sum_{i=1}^L \left( {\bf I} - {\bf M}\right)^{-1}_{iw} ~,
\label{UpperBound}
\end{equation}
where the matrix indices run from Hamming distance $1$ through $L$. $\tau_1$
determines an upper bound for the entire population's barrier crossing time.
For parameter settings in the fixation regime, where $M L \mu \ll 1$, sampling
fluctuations cause the population to converge onto $M$ copies of a single
genotype. As is well known from the theory of neutral evolution 
\cite{Kimurabook}, this set of identical genotypes performs a random walk
through the genotype space at the same rate as a single
individual. Thus, in this limit, $\tau_1$ gives a reasonable
prediction for the entropy-barrier crossing time. However, for
Fig. \ref{theo+exp_fig}(a)'s parameter settings ($M=250$, $\mu
=0.005$, and $L=10$) that give $M L \mu = 12.5$, we find that $\tau_1
\approx 23000$, almost independent of valley width $w$. Of course,
this grossly overestimates the observed barrier crossing times, which
vary from $\BCTime \approx 25$ for $w=2$ to $\BCTime \approx 227$ for
$w=4$.

For $M$ independent random walkers, one might simply assume that the
waiting time would be roughly a factor $M$ slower, i.e.
$\tau_M = \tau_1/M$. Unfortunately, this leads to $\tau_M \approx 93$
which overestimates the observed time for $w=2$ and underestimates
$\BCTime$ for $w=4$.

The precise probability $\bar{p}_w(t)$, that {\em none} of $M$
independent random walkers starting at a Hamming distance $w$ have
found the portal by time $t$, is given by:
\begin{equation}
\label{M_walkers_not_found}
\bar{p}_w(t) = \left(\sum_{i=1}^L \MuOp^t_{iw} \right)^M ~.
\end{equation}
From this, one estimates the average entropy-barrier crossing time $\tau$ to be:
\begin{eqnarray}
\nonumber
\tau_M & = & \sum_{t=1}^{\infty} t \left[ \bar{p}_w(t-1)-\bar{p}_w(t) \right] \\
  & = & \sum_{t=0}^{\infty} \bar{p}_w(t) ~.
\label{M_walkers_av_time}
\end{eqnarray}
For Fig. \ref{theo+exp_fig}(a)'s parameters,
Eq. (\ref{M_walkers_av_time}) gives $\tau \approx 15$, $58$, and $117$
for barrier widths $w=2$, $3$, and $4$, respectively. These values
underestimate each observed waiting time by almost a factor of
$2$. Apparently, sampling fluctuations cause the population to explore
the genotype space less rapidly than {\em independent} random walkers
do. As already noted above, the reason for this is that sampling
convergence leads different individuals to evolve genetic correlations
to some degree.

One way to think about this is to investigate genealogies. Ref.
\cite{Derrida&Peliti} showed that the probability ${\rm Pr}(t)$ for
two randomly chosen individuals in the current population to have had
a common ancestor less than $t$ generations ago is approximately
given by
\begin{equation}
{\rm Pr}(t) \approx 1-e^{-t/M} ~.
\end{equation}
This means that, on
average, a pair of individuals has only undergone $M L \mu$ mutations
each since the time $t \approx M$ they descended from a common
ancestor.  When $M\mu$ is not much larger than $1$, this implies that
two individuals are more strongly correlated genetically than random
genotypes. Due to this, it is easy to see, at least qualitatively,
that the entropy-barrier crossing time is longer than that predicted
for independent random walkers. The correlation, or clustering, of
individuals in genotype space leads the population to explore the
valley's neutral volume at a slower rate. Thus, the predictions obtained
by assuming $M$ random walkers, as given by Eqs.
(\ref{M_walkers_not_found}) and (\ref{M_walkers_av_time}), are lower
bounds to the actual waiting times.

It turns out that the upper (Eq. (\ref{UpperBound})) and lower
(Eq. (\ref{M_walkers_av_time})) estimates do not tightly bound the
actual waiting times $\BCTime$. They may differ by several orders of
magnitude. Fortunately, the lower bound obtained from
Eqs. (\ref{M_walkers_not_found}) and (\ref{M_walkers_av_time})
typically produces reasonable order-of-magnitude estimates for
parameter regimes in which $M L \mu > 1$. This order-of-magnitude
estimate gives the following scaling relation for the entropy-barrier
crossing time
\begin{equation}
\tau \approx \frac{2^L}{M L \mu} ~.
\label{entropic_scaling}
\end{equation}

\subsection{Anomalous Scaling}

The order-of-magnitude estimate given by Eq. (\ref{entropic_scaling})
predicts that the the entropy-barrier crossing time $\tau$ scales
inversely with both $\mu$ and $M$. This scaling is, of course, exactly
one's intuitive expectation: the rate at which the genotype space is
explored is proportional to both mutation rate $\mu$ and population
size $M$. $M$ individuals cover $M$ times as much genotypic ``ground''
as one individual. Individuals that ``move'' twice as fast, cover
twice as much ground as well. And so, the waiting time should be
inversely proportional to both $M$ and $\mu$, which set the
exploration rate.

\begin{figure}[htbp]
\centerline{\epsfig{file=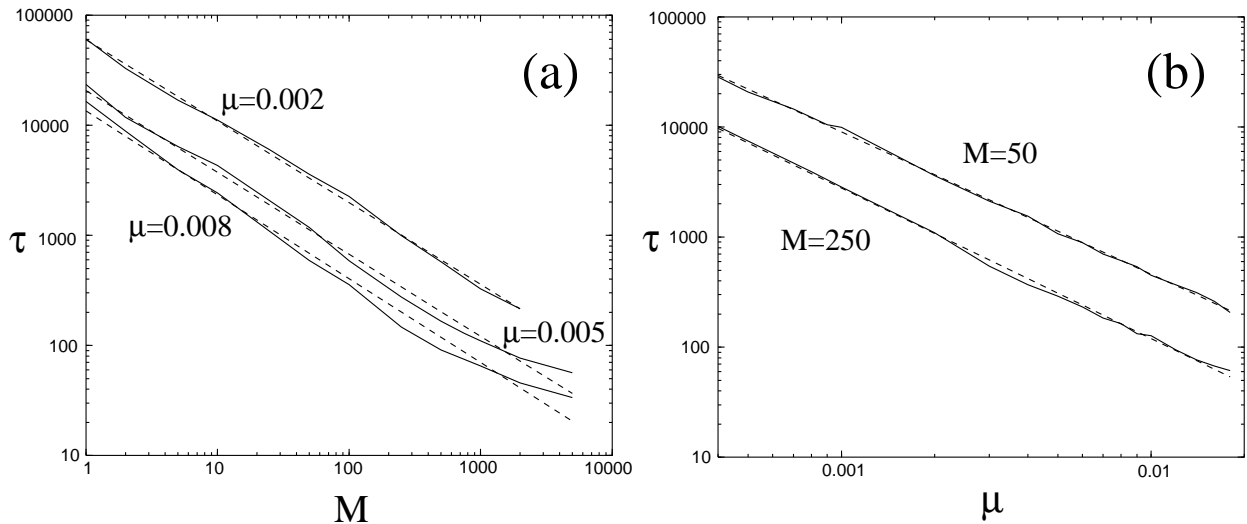,width=6.5in}}
\caption{Entropy-barrier crossing time $\tau$ as a function of population
size $M$ and mutation rate $\mu$. The solid lines are data obtained from
simulations, while the dashed lines show the estimated scaling functions.
All experiments were done with genotypes of length $L =10$ and a barrier
width of $w=5$. All axes are shown on logarithmic scales. In Fig. (a) $\tau$
is shown as a function of population size $M$ for three values of the mutation
rate $\mu$. In Fig. (b) $\tau$ is shown as a function of the mutation rate $\mu$
for two values of population size $M$. The approximately straight lines show
that $\tau$ scales as a power law in $M$ when $\mu$ is kept constant and,
vice versa, as a power law in $\mu$ when $M$ is kept constant. Table
\ref{exponent_table} lists the estimated exponents for the power laws.
These were used to plot the dashed lines.}
\label{neut_scaling}
\end{figure}

In light of this, it is interesting that data from simulations shows
that the entropy-barrier crossing time $\tau$ scales as a power law
in both $\mu$ and $M$, but {\em not} with exponents equal to $-1$,
as the preceding simple argument suggests. To be clearer on this
point, Fig. \ref{neut_scaling} illustrates the observed scaling
behavior of the entropy-barrier crossing time as a function of $M$
and $\mu$.

The solid lines plot the data obtained from simulations while the
dashed lines show scaling (power-law) functions that were fitted to
the experimental data. All axes use logarithmic scales. All
simulations were performed with genotypes of length $L=10$ bits.  In
all of the runs, at time $t=0$ all individuals start at Hamming
distance $w=5$ from the portal. Figure
\ref{neut_scaling}(a) shows $\tau$'s dependence on $M$ for three
different values of $\mu$. The approximately straight lines show that
the entropy-barrier crossing time depends roughly as a power law on
the population size $M$:
\begin{equation}
\label{def_alpha}
\tau \propto \frac{1}{M^{\alpha}} ~.
\end{equation}
Of course, the scaling exponent $\alpha$ may itself depend on $\mu$.

Similarly, Fig. \ref{neut_scaling}(b) shows the dependence of $\tau$
on $\mu$ for two different values of $M$. In this case too, the curves
appear well approximated by a straight line, indicating that for fixed
$M$ the dependence on $\mu$ is roughly given by
\begin{equation}
\label{def_beta}
\tau \propto \frac{1}{\mu^{\beta}} ~,
\end{equation}
where $\beta$ may again depend on the population size $M$. In Table
\ref{exponent_table} the exponents of the estimated dashed lines in Figs.
\ref{neut_scaling}(a) and \ref{neut_scaling}(b) are given, along with their
estimated errors.
\begin{table}[htbp]
\begin{center}
\begin{tabular}{|l|l|l|l|}
$\mu$ & $0.002$ & $0.005$ & $0.008$ \\ 
\hline
$\alpha$ & $0.740 \pm 0.01$ & $0.744 \pm 0.02$ & $0.761 \pm 0.03$ \\
\hline
\hline
M & $50$ & $250$ &  \\
\hline
$\beta$ & $1.292 \pm 0.008$ & $1.365 \pm 0.014$ &  \\
\end{tabular}
\end{center}
\caption{Estimated exponents $\alpha$ and $\beta$ as defined by Eqs.
(\ref{def_alpha}) and (\ref{def_beta}).}
\label{exponent_table}
\end{table}

\end{multicols}

\begin{multicols}{2}

The values of the exponents $\alpha$ for different $\mu$ and $\beta$
for different $M$ are very close to each other; $\alpha \approx 3/4$
and $\beta \approx 4/3$. It is clear, however, that they are not {\em
constants}: $\alpha$ does depend on $\mu$ and $\beta$ on $M$.  Note
that the estimates for $\alpha$ are all below $1$, while those for
$\beta$ are above $1$. Thus, doubling the population size decreases
$\tau$ by {\em less} than a factor of two, while doubling the mutation
rate decreases $\tau$ with {\em more} than a factor of
two. Intuitively, what is happening is that, due to the clustering in
the population, doubling the population size does not lead to a
doubling of the exploration rate.  That is, some of the ``added''
members in the larger population will simply occur at genotypes where
other members of the population are already located.  Thus, they do
not contribute to additional novel exploration. In contrast, doubling
the mutation rate not only doubles the rate of movement (diffusion) of
individuals in the population, it additionally decreases the
clustering and so reduces genetic correlations. Due to the combination
of these two effects, the entropy-barrier crossing time decreases more
than a factor two.

Of course, one would like to predict these anomalous exponents. In
principle, they should be calculable from knowledge of the clustering
structure of the population at different values of $M$ and $\mu$. For
example, if we view the population as a blob or collection of blobs in
genotype space, one would like to know how many distinct genotypes, on
average, are neighboring one or more individuals of the
population. Roughly speaking, we would like to know the average
``surface area'' of the population in genotype space. Knowledge of
this statistic would then supply us with the average probability that
a mutation leads to a genotype not currently present in the population.
This, in turn, quantifies the population's rate of exploring novel
genotypes, while taking into account genetic correlations. Although
several statistics of these genotype blobs were calculated in
Ref. \cite{Derrida&Peliti}, we have at present not been able to adapt
these results to infer the necessary type of statistics just
outlined. The analytical prediction of the scaling exponents $\alpha$
and $\beta$ thus awaits further progress. For the present, we will use
our order-of-magnitude estimate in Eq. (\ref{entropic_scaling}) to
compare the entropy-barrier crossing time with the fitness-barrier
crossing times.

In summary, we analyzed the fitness- and entropy-barrier crossing
times for the simplest (single peak) landscapes in which both types of
barrier occur. Our results are summarized by the scaling relations
of Eqs. (\ref{empirical_scaling}) and (\ref{entropic_scaling}). In the
following sections, we apply the preceding analysis to more
complicated fitness functions that contain multiple fitness and
entropy barriers.

\section{Traversing Complex Fitness Functions}
\label{royal_staircase}

Up to this point, to make analytical progress we focused on fitness
functions that were intentionally simple: a single portal and a single
peak in genotype space. Despite this, the analysis of barrier crossing
times just developed can be extended with relative ease to more
complicated evolutionary processes. To illustrate this extension of the
theory, we now introduce a class of more complicated fitness functions
that contain multiple fitness and entropy barriers of tunable width and
height.  That is, in this class of fitness functions, the population
may have to cross both a fitness {\em and} entropy barrier to escape
from its metastable state. Since the relative sizes of both these
types of barriers can be tuned, we can explicitly compare the time
scales for crossing fitness and entropy barriers within the same
evolutionary process. 

\subsection{The Royal Staircase with Ditches}

The class of fitness functions which we call the {\em Royal Staircase
with Ditches} is closely related to the {\em Royal Staircase} and
{\em Royal Road} fitness functions that we have analyzed previously
\cite{Crut99a,Nimw98b,Nimw98a,Nimw97a,Nimw97b}. Those fitness functions
did not contain fitness barriers but instead consisted of a series of
entropy barriers. The function class of Royal Staircases with Ditches
generalizes these fitness functions and is defined as follows:

\begin{itemize}
\item{Genotypes consist of bit sequences of length $L$, interpreted as
$N$ blocks of $K$ bits each: $L=NK$.}

\item{The blocks are ordered, not in the sense that they correspond to 
particular positions in the genotype, but only in the sense that they
are indexed $1$ through $N$. Note that since our evolutionary process
does not include recombination, the population dynamics is invariant under
arbitrary permutations of a genotype's bits. For convenience, we order
the blocks from left to right in the genotypes. That is, bits $1$
through $K$ belong to the first block, bits $K+1$ through $2 K$ belong 
to the second block, and so on.}

\item{The $2^K$ possible configurations of the $K$-bit blocks each 
are divided into three classes.}
\begin{enumerate}
  \item{Type-$A$ blocks consist of a configuration with $K$ ones:
  \begin{equation}
  A = \underbrace{111 \cdots 111}_{K} ~.
  \end{equation}
  }
  \item{Type-$B$ blocks consist of a configuration with $K-w$ ones
  and $w$ zeros:
  \begin{equation}
  B = \underbrace{111 \cdots 111}_{K-w} \overbrace{000 \cdots 000}^w ~.
  \end{equation}
  As will become clear, the parameter $w$ controls the {\em width} of the barriers.}
  \item{All other $2^K-2$ configurations are denoted as Type-$*$ blocks.}
\end{enumerate}

\item{A genotype $s$ with blocks $1$ through $n-2$ of type $B$ and block $n-1$
of type $A$ receives fitness $f(s) = n$. These genotypes have the structure:
\begin{equation}
\underbrace{BB \cdots BB}_{n-2} A \overbrace{**\cdots**}^{N-n+1} ~.
\end{equation}
Note that the configurations of blocks $n$ through $N$ are immaterial (denoted
$*$) when the first $n-1$ blocks occur in the above genotype configuration.}
\item{Genotypes $s$ with blocks $1$ through $n-2$ of type $B$ , block $n-1$ of
type $*$, and block $n$ of type $A$ receive fitness $f(s) = n-h$. These
genotypes have the structure:
\begin{equation}
\underbrace{BB \cdots BB}_{n-2} *A \overbrace{**\cdots**}^{N-n} ~,
\end{equation}
Again, the configurations of blocks $n+1$ through $N$ are immaterial. The
parameter $h$ controls the {\em height} of the fitness barrier.}
\end{itemize}
The Royal Staircase fitness functions that we studied in Refs. \cite{Nimw98a}
and \cite{Nimw98b} are a special case ($w=0$) of the Royal Staircase with
Ditches class of fitness functions. For the special case $w=0$ there are no
fitness barriers and a genotype has fitness $n$ when the first $n-1$ blocks
are $A=B$ (all $1$s) types and the $n$th block is set to any of the $2^K-1$
other configurations.

Setting $w=1$ produces a somewhat degenerate case that we will not consider. 

For values of $w \geq 2$, there is a genuine fitness barrier of width
$w$ bits. For instance, consider the case where, at some point in
time, the highest fitness in the population is $f = 4$. This
corresponds to genotypes that have first and second blocks of type $B$
and a third block of type $A$. In this case, a {\em portal} genotype
$\Omega_n$, corresponding to a fitness of $5$, is obtained when the
fourth block is set to type $A$ and the third block is changed from
type $A$ to type $B$. Genotypes with fitness $4$ can mutate their
fourth block, until it becomes type $A$, without changing their
fitness. That is, the fourth block may be changed into type $A$ along
a {\em neutral path} and setting the $4$th block correctly corresponds
to crossing an entropy barrier. However, after that, the third block
needs to be changed from type $A$ to type $B$. All intermediate type
$*$ blocks give genotypes a reduced fitness of $f = 4-h$.  We call
these {\em ditch} genotypes, since they are located in a lower-fitness
region in genotype space that separates genotypes with fitness $4$
from genotypes with fitness $5$.

\subsection{Evolutionary Dynamics}

We evolve populations on the Royal Staircase with Ditches under a simple
selection and mutation dynamics similar to the one outlined in Sec.
\ref{evolutionary_dynamics}. This consists of the following steps.
\begin{enumerate}
\item{At time $t=0$ a population of $M$ random binary-allele genotypes
(bit sequences) of length $L$ is created. These $M$ individuals
constitute the initial population.}
\item{The fitness of all $M$ individuals is determined, using the
function defined in the previous section.}
\item{$M$ individuals are sampled from the
population, with replacement, and with probability proportional to
their fitness. That is, the population undergoes fitness-proportional
selection in discrete generations.}
\item{Each bit in each of the $M$ selected individuals is mutated with 
a probability $\mu$. The $M$ individuals thus obtained form the new
generation.}
\item{The procedure is repeated from Step 2.}
\end{enumerate}
We evolve the population according to the above dynamics until genotypes
of optimal fitness have been discovered and the population appears to
have reached a stable average fitness. During each run, we estimate
a number of statistics---such as, the average time until individuals
of a certain fitness appear for the first time.

\subsection{Observed Population Dynamics}

\begin{figure}[htbp]
\centerline{\epsfig{file=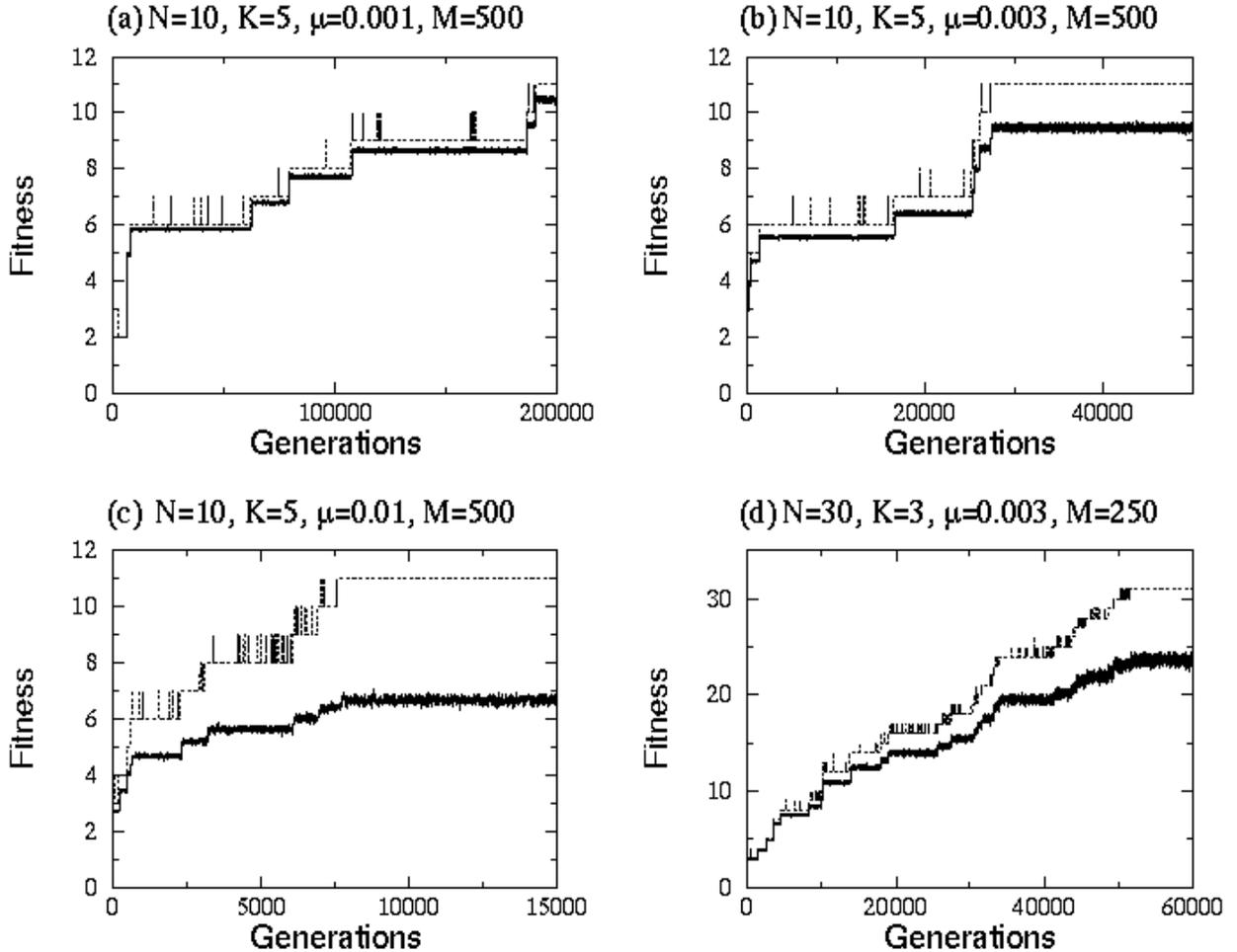,width=6.5in}}
\caption{Four runs of the Royal Road with Ditches population dynamics
with four different parameter settings. The upper dashed lines plot
best fitness and the lower solid lines, average fitness $\AvFit$ in
the population as a function of time, measured in generations. The
parameter settings for each run are indicated above each figure. All
fitness functions have barriers with a width of $w=2$ and a height of
$h=1$. The values of the average fitnesses were obtained by taking a
running average over $10$ generations. This reduces the relatively
large fluctuations in average fitness between successive generations.}
\label{single_runs}
\end{figure}

The population dynamics under Royal Staircase with Ditches functions
is qualitatively very similar to that under the Royal Road and Royal
Staircase fitness functions. Samples of this typical behavior are
shown in Figs. \ref{single_runs}(a)-(d). The plots there show the
average $\AvFit$ (lower, solid lines) and best (upper, dashed lines)
fitnesses in the population over time for four single runs with four
different parameter settings. The parameter settings for each run are
indicated above each figure, except for the barrier widths $w$ and
barrier heights $h$ that were used. All runs used barriers of widths
$w=2$ and heights $h=1$.
\end{multicols}


\begin{multicols}{2}

The qualitative dynamics follows the typical alternation of long periods
of stasis ({\em epochs}) in the average population fitness and short bursts
of {\em innovation} to higher average fitness; a class of evolutionary
dynamics that we call {\em epochal evolution} \cite{Nimw97a}. At the
beginning of a run the average and best fitness are low. This simply
reflects the fact that high-fitness genotypes are very rare in genotype
space and therefore do not occur in an initial random population. A balance
is quickly established between selection and mutation that leads to a
roughly constant average fitness in the population. This period of stasis
we call an {\em epoch}. After some time, a mutant may cross the fitness
barrier and discover a {\em portal}, i.e., a higher-fitness genotype.
Relatively frequently, this high-fitness mutant is lost through sampling
fluctuations or deleterious mutations. Such events are seen as isolated
spikes in the best fitness in Fig. \ref{single_runs}.  Eventually, one
of these beneficial mutants spreads through the population---an
{\em innovation} occurs. At this point the average fitness increases,
until a new equilibrium between selection and mutation is established.

Although many properties of epochal evolution can be treated
analytically \cite{Nimw97b}, here we will focus solely on the {\em
epoch times}. These are the average times between the start of a given
epoch and the start of the next. Average epoch times can be obtained
from simulation data by tracing backward in time the behavior of the
population's best fitness. The population dynamics runs until genotypes
of the highest possible fitness $N+1$ have established themselves in the
population for an extended period of time. (For certain parameter
settings---specifically, those beyond the error threshold---genotypes
with the highest fitness may not stabilize in the population at
all. For such parameter settings, an epoch time can be effectively
infinite. We will not consider such parameter settings explicitly
here.) From there we trace backwards the {\em last} time $t_n$ that
fitness $n$ was the highest in the population, for all values of
$n$. The differences $t_n-t_{n-1}$ give the epoch times $\tau_n$.
Some fitness levels may not occur during a run. For instance, in Fig.
\ref{single_runs}(a) fitness $f = 1$ never occurs, since the population starts
out with genotypes of fitness $2$. The epoch time $\tau_1$ is therefore $0$ for
this particular run. Average epoch times are then estimated by averaging the
epoch time $\tau_n$ over an ensemble of runs. In the following we calculate
analytical approximations to these epoch times $\tau_n$, using the
results from the preceding development.

\subsection{Epoch Quasispecies and The Statistical Dynamics Approach} 

In order to approximate epoch times analytically, we need to determine
the average proportion of the population at highest fitness during
each epoch. This is the equivalent of $P_{\Pi}$ in Sec.
\ref{crossing_single_barrier}. From this we can estimate the rate of
creation of genealogies in the {\em ditch} between the neutral networks
of two successive epochs. Then we need to calculate
the average population fitness during each epoch to determine the
average life time of these ditch genealogies. For the fitness
functions studied in the Sec. \ref{crossing_single_barrier} these
quantities were relatively straightforward to calculate. There, individuals
were either on the peak or in the valley, at a certain distance from
the portal. For the Royal Staircase with Ditches fitness functions the
situation is more complicated.

In principle, one can calculate the current equivalent of
$P_{\Pi}$ by representing the population as a distribution of {\em
genotypes} and calculating metastable genotype distributions for each
epoch. This is typically done in population genetics models
\cite{Ewensbook} and in the standard quasispecies models of molecular
evolution \cite{Eigen&McCaskill&Schuster89}. However, since genotype
spaces are typically very large, an analytical treatment that
explicitly takes into account finite-population effects is generally
infeasible within this genotypic representation. To address this
problem, we introduced an alternative approach that we call {\em
statistical dynamics}
\cite{Crut99a,Nimw97a,Nimw97b}. There one chooses a relatively small number of
{\em macroscopic variables} with which to describe the population at any
given time. Other degrees of freedom are then averaged out using a
{\em maximum entropy} method similar to the Gibbs method from statistical
mechanics. We will use this approach below and simply refer the reader to
Refs. \cite{Crut99a} and \cite{Nimw97b} for more extensive treatment of
statistical dynamics and a discussion of the relation of this approach to
standard quasispecies
theory, mathematical population genetics, and other theories from the field
of evolutionary computation \cite{PrugelBennett&Shapiro94}.

We represent the population state at any point in time by a {\em
fitness distribution}. That is, instead of describing the population
by the relative frequencies of all genotypes in the population, we
describe it by the relative frequencies of different fitness values.
Of course, a given fitness distribution does not uniquely specify a
population of genotypes. In order to construct the dynamics on the
level of fitness distributions, we must ``average out'' the additional
genotypic degrees of freedom somehow. We do this by assuming, at each
generation, a maximum entropy distribution of genotypes given
the distribution of fitness. In the Royal Staircase with
Ditches fitness functions, this translates into assuming that each
string with a given fitness $f$ is equally likely to be {\em any} of
the genotypes with fitness $f$. That is, a genotype with fitness $n$
will have its first $n-1$ blocks each in a specific state. Blocks $n$
through $N$ are assumed to occur in any of their $2^K$ possible states
with equal probability. Similarly, for a ``ditch'' genotype with
fitness $n-h$, there are $n-1$ blocks in fixed genotypic states and
one block of type $*$.  We assume this $*$-block occurs in any of its
$2^K-2$ possible bit configurations with equal probability. Similarly,
we assume that blocks $n+1$ through $N$ are equally likely to be in
any of their $2^K$ configurations. That is, we assign equal
probability to all genotype distributions that are consistent with the
given fitness distribution.  Since we know the dynamics on the
genotype level, we can construct the expected dynamics on the level of
fitness distributions under these assumptions.

Formally, we represent the population at time $t$ as a vector $\vec{P}(t)$
with components $P_n(t), n = 1,2, \ldots, N+1$ that denote the proportion
of fitness-$n$ individuals in the population and with components $P_{n,*}(t)$
that denote the proportion of fitness $n-h$ individuals in the ditch
between fitness $n$ and fitness $n+1$ genotypes. (Note that in the
cases where $h$ is an integer the distribution $\vec{P}(t)$ is not simply
a fitness distribution, since it distinguishes ditch individuals from nonditch
individuals at the same fitness.)

We then construct a generation operator ${\bf G}$, similar to the one
in section \ref{metastable_quasispecies}, that acts on a fitness
distribution $\vec{P}(t)$ and returns the {\em expected} fitness
distribution $\langle \vec{P}(t+1) \rangle$ at the next generation.
That is, the dynamics at the level of fitness is to be given by
\begin{equation}
\label{formal_one_gen_dyn}
\langle \vec{P}(t+1) \rangle
  = \frac{{\bf G} \cdot \vec{P}(t)}{\AvFit} ~, 
\end{equation}
where $\AvFit$ is the average fitness in the population.

During epoch $n$, in which the best fitness occurring in the
population is $n$, there will be a roughly constant fitness
distribution $\vec{P}^n$. These vectors $\vec{P}^n$ are the solutions
of the fixed point equations $\langle \vec{P}(t+1) \rangle = \vec{P}(t)$
determined by Eq. (\ref{formal_one_gen_dyn}). Once the operator ${\bf
G}$ is constructed, the epoch distribution can be calculated quite
easily. More specifically, in Refs. \cite{Nimw97a} and \cite{Nimw97b}
we showed that the metastable fitness distribution that occurs during
epoch $n$ is determined by projecting the operator ${\bf G}$ onto all
dimensions with fitness smaller than or equal to $n$ and calculating
the principal eigenvector of this projected operator. Determining the
epoch-$n$ quasispecies reduces, in this way, to finding the principal
eigenvectors of the matrix ${\bf G}$ restricted to components with
fitness lower than or equal to $n$. In App. \ref{quasispecies_calc}
the generation operator ${\bf G}$ for the Royal Staircase with Ditches 
is constructed explicitly, and analytical approximations to the
epoch quasispecies distributions $\vec{P}^n$ are calculated as well.
Since the expressions we find for the fitness distributions
$\vec{P}^n$ are rather cumbersome and we will not give them
here. 

\subsection{Crossing the Fitness Barrier}
\label{crossing_fit_barrier}

With the analytical expressions for the epoch fitness distributions
$\vec{P}^n$ in hand, we can calculate the expected epoch times $\tau_n$.
An epoch ends via an innovation---a process that proceeds in two stages.
First, a portal genotype of fitness $n+1$ is created. And second, this
beneficial mutant spreads through the population, rather than being lost.

In order to calculate the average time until a portal genotype is
discovered we calculate the probability $P^{\rm seed}$ that a {\em
single selection plus mutation} from the current population seeds
a new lineage that discovers the portal. That is, either by creating a
new lineage in the ditch that discovers the portal or else by
producing a jump mutation that becomes a portal genotype at once.

This calculation is very similar to that in
Sec. \ref{crossing_single_barrier}. First of all, the portal is
unlikely to be discovered by anything other than either a jump
mutation from a fitness-$n$ individual or by a mutation from a ditch
genotype.  Moreover, ditch lineages are very unlikely to be seeded by
anything other than mutants of fitness-$n$ genotypes. Therefore, we
can write $P^{\rm seed}$ as
\begin{eqnarray}
\nonumber
P^{\rm seed} & = & \frac{n P^n_n (1-\mu)^{(n-2) K}}{f_n 2^K} \\
  & \times &
  \sum_{i=0}^K \epsilon_i \left( \MuOp_{iw} - (1-\mu)^K \delta_{iw} \right) ~.
\label{formal_p_seed}
\end{eqnarray}
The first factor $n P^n_n/f_n$ gives the probability that a fitness
$n$ individual is selected. For the offspring of this individual to
end up in the ditch, its $n$th block should be type $A$
(thereby contributing the factor $2^{-K}$) and its first $n-2$ blocks
should be left undisturbed by mutation (corresponding to the factor
$(1-\mu)^{(n-2)K}$). The terms within the sum give the probability
that a valley lineage is seeded by mutation of the $(n-1)$st block at
Hamming distance $i$ from the portal. Finally, the factors
$\epsilon_i$ give the probabilities that a lineage, starting at
Hamming distance $i$ from the portal, discovers the portal. They are
analogous to the $\epsilon_i$ of Sec. \ref{valley_lineages}. Note that
the term $\epsilon_0 \MuOp_{0w} = 1 \cdot
\mu^w (1-\mu)^{K-w}$ corresponds to a jump mutation from a fitness-$n$
genotype directly to a portal configuration. Again, changes to
blocks $n+1$ through $N$ are immaterial. 

The $\epsilon_i$ are calculated paralleling the development in Sec.
\ref{valley_lineages}. The only differences are 
that the genotype length $L$ is now the block length $K$ here, since
we are only interested in mutations in the $(n-1)$st block, and that
the average number of offspring is slightly modified. The average
number of {\em ditch} offspring $r$ that an individual in the ditch
produces on average is given by
\begin{equation}
r = \frac{(n-d) (1-\mu)^{(n-1)K}}{f_n} = \frac{n-d}{n} ~.
\end{equation}
The expressions for the $\epsilon_i$ in this case can be obtained by
replacing the factor $1/\AvFit$ in the equations in Sec.
\ref{valley_lineages} by $r$.

Eq. (\ref{formal_p_seed}) can be
further simplified and we eventually obtain the result that
\begin{eqnarray}
P^{\rm seed} & = & \\ \nonumber
  & - & \frac{P^n_n}{2^K} \left[ \epsilon_w +
\frac{n}{(1-\mu)^K (n-d)} \log(1-\epsilon_w) \right] ~,
\label{simpler_p_seed}
\end{eqnarray}
where $\epsilon_w$ is given by Eqs. (\ref{EpsilonFormula}) and
(\ref{EpsilonFormulaII}) and $P^n_n$ is determined as outlined in App.
\ref{quasispecies_calc}.

Thus, once we have calculated $\epsilon_w$ using the results of
Sec. \ref{valley_lineages} and calculated $P^n_n$ using the derivation
described in App. \ref{quasispecies_calc}, we find that the average
number $g_n$ of epoch-$n$ generations until a genotype of fitness
$n+1$ is created is given by
\begin{equation}
g_n = \frac{1}{1- (1-P^{\rm seed})^M}.
\label{discover_time}
\end{equation}
In this, we have neglected the correction term $\langle dt \rangle$
of Sec. \ref{LengthValleyBushes} for the time between the seeding of the
ditch lineage that leads to the discovery of the portal and the actual
time at which the portal is discovered. 

We now must calculate the second stage of epoch termination; namely,
the probability $\pi_n$ that a newly discovered mutant of fitness
$n+1$ spreads through the population. In Ref. \cite{Nimw97b} we showed
how a diffusion equation approach \cite{Kimura62,Kimura64} can be used
to estimate this probability. Applying this here, the probability
$\pi_n$ that a mutant of fitness $n+1$ will spread through the
population is given by:
\begin{equation}
\label{spread_prob}
\pi_n = 1 -\exp \left( -2 \frac{(n+1) (1-\mu)^K}{n} +2 \right) ~.
\end{equation}
A mutant must be discovered $1/\pi_n$ times on average before it stabilizes
and spreads through the population. Thus, $g_n/\pi_n$ gives the average 
number of generations until a portal genotype is discovered that
spreads through the population---an innovation.

Finally, we have to account for the possibility that epoch $n$ does
not occur at all during a run. Only a fraction $P_e(n)$ of the runs
contain epoch $n$, since, if a higher-fitness genotype occurs in the
initial random population, epoch $n$ will be skipped. The proportion
$P_e(n)$ is simply the probability that {\em no} genotypes of fitness
greater than $n$ occur in the initial population. This is given by
\begin{equation}
P_e(n) = \left[ \sum_{i=0}^{n-1} \frac{1}{2^{K i}} \left(1-\frac{1}{2^K}
\right) \right]^M ~.
\label{ProbNoEpochN}
\end{equation}
Putting Eqs. (\ref{discover_time}), (\ref{spread_prob}), and
(\ref{ProbNoEpochN}) together, the theoretical predictions for the epoch
times $\tau_n$ become
\begin{equation}
\tau_n = P_e(n) \frac{g_n}{\pi_n} ~.
\label{ep_times}
\end{equation}

\subsection{Theoretical and Experimental Epoch Times}
\label{theo_exp_ep_times}

\begin{figure}[htbp]
\centerline{\epsfig{file=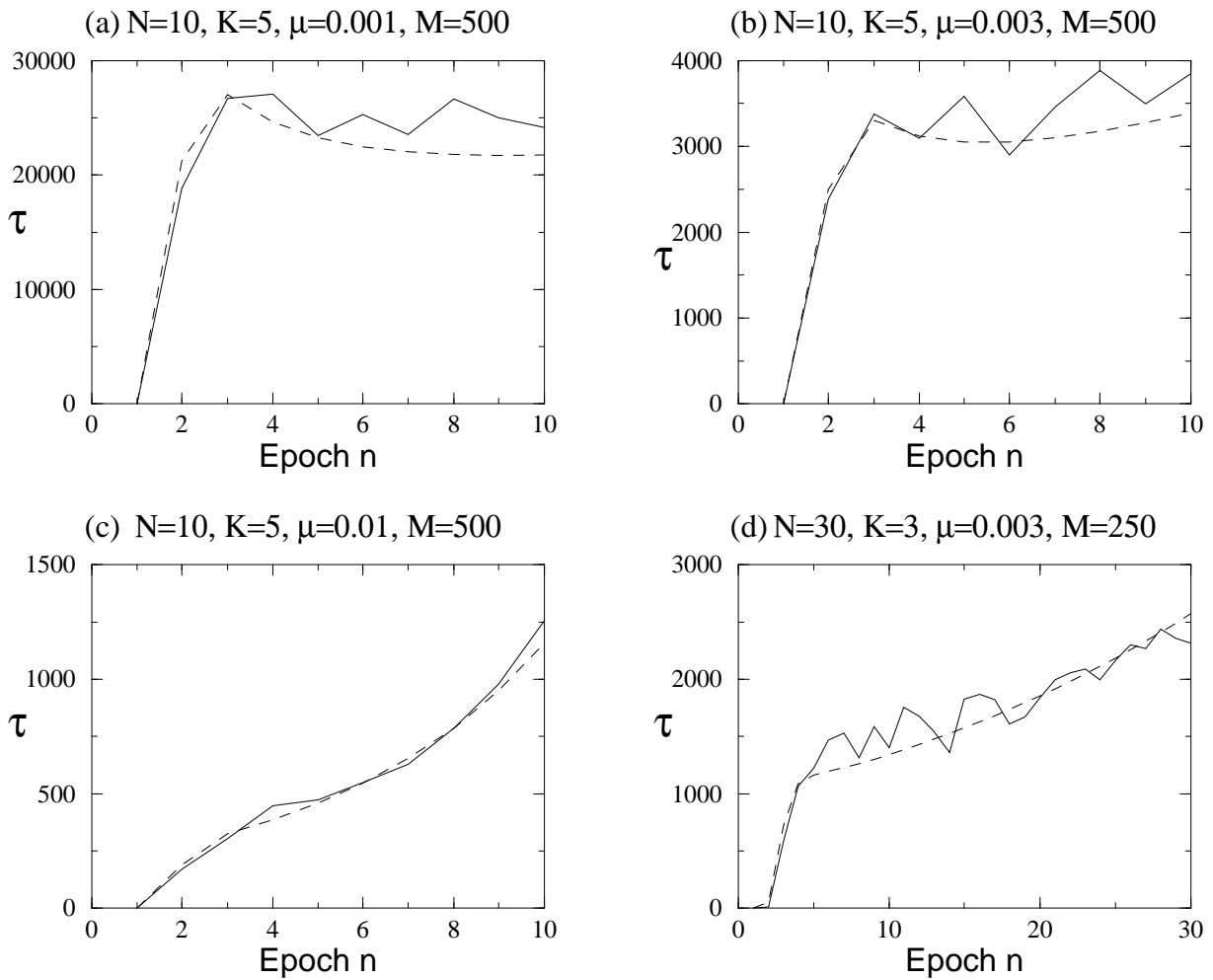,width=6.5in}}
\caption{Experimentally estimated (solid lines) and theoretically predicted
(dashed lines) epoch times $\tau_n$ for the four different parameter settings
of Fig. \ref{single_runs} as a function of epoch number $n$. All fitness
functions have ditches of width $w=2$ and height $h=1$. The experimental epoch
times are an average over $150$ runs for Fig. (a) and over $250$ runs for Figs.
(b), (c), and (d).}
\label{epoch_times}
\end{figure}

We tested the predictions of Eq. (\ref{ep_times}) against experimentally
obtained average epoch times for the same parameter settings used in Fig.
\ref{single_runs}. The results are shown in Fig. \ref{epoch_times}. Epoch
times $\tau_n$ are shown as a function of epoch number $n$. The dashed lines
give the theoretical predictions of Eq. (\ref{ep_times}) and the solid lines,
the experimentally estimated averages. The data in Fig. \ref{epoch_times}(a)
is an average over $150$ evolutionary runs. All other plots are averages over
$250$ runs. The fluctuations in the experimental lines indicate that there is 
a large variance of epoch times between runs and that, therefore, the raw data
is rather noisy. Despite this, the figure demonstrates that the theory
estimates epoch times quite accurately. 

First, these results show that the analysis presented in Sec.
\ref{crossing_single_barrier} can be usefully applied to more complicated
fitness functions that posses many fitness and entropy barriers. For
simplicity in the present case, we restricted ourselves to fitness
functions with barriers of equal height and width. However, since our
statistical dynamics methods analyze epochs in a piecewise manner, the
theoretical results easily extend to more general cases with barriers
of variable width and height. The essential ingredient of the analysis
is still the genealogy statistics of valley lineages in the ditch that
connects to portal genotypes. The only additional ingredients required
by the analysis for more complicated cases are (i) the rate of
creation of new lineages in the ditch and (ii) the distribution of
Hamming distances to the portal at which these lineages are seeded. In
the case of the Royal Staircase with Ditches, this was largely
determined by the proportion $P^n_n$ of fitness $n$ individuals during
each epoch $n$. Apart from this determining factor, the actual
crossing of fitness barriers is still governed by the scaling
relations presented in Sec. \ref{scaling_bar_cross_time}. In
particular, the qualitative remarks at the end of
Sec. \ref{scaling_bar_cross_time} still hold in these more general
settings. Epoch times grow very rapidly with barrier width and quite
slowly with barrier height.

\end{multicols}


\begin{multicols}{2}

Second, and perhaps of more immediate interest, the general shape of the
curves in Fig. \ref{epoch_times} reveals several novel population dynamical
phenomena. For the lower mutation-rate runs (Figs. \ref{epoch_times}(a),
\ref{epoch_times}(b), and \ref{epoch_times}(d)), the epoch times show a
distinct break between the relatively small $\tau_1$ and the much larger
$\tau_2$. This jump simply reflects the fact that for these population sizes,
it is very likely that individuals with fitness $2$ occur in the initial
random population. Due to this, epoch $1$ is skipped most of the time.
However, individuals of fitness $3$ are unlikely to occur in the initial
population---the occurrence of a fitness $3$ individual is roughly one in
$10^3$---so that the second epoch is not skipped. In Fig. \ref{epoch_times}(a)
and \ref{epoch_times}(b), the epoch time curve reaches a maximum after this
quick jump and then drops off slightly. In Fig. \ref{epoch_times}(a) with
mutation rate $\mu = 0.001$ the times seem to just reach a minimum around
the last epoch, number $10$. The curves in Fig. \ref{epoch_times}(b) reach
a minimum epoch time somewhere around epoch $5$ and then start increasing
again. The behavior typically occurs for a range of parameter values with
low mutation rates and with block sizes $K$ that are not too small. For
instance, the theoretical analysis predicts that if there were more than
$10$ blocks in the case of Fig. \ref{epoch_times}(a), then the curve would
start to rise after epoch $10$.

Qualitatively, this behavior can be understood as follows. Since the
barriers all have equal {\em height} $h$, the barriers at later epochs
are relatively more shallow than those visited during early epochs.
From the analysis in the last section recall that the average number
of offspring that ditch individuals produce in the ditch is $r =
(n-h)/n$. As $n$ increases, this number approaches $1$ from
below. That is, ditch lineages survive longer for later epochs. This
effect, of course, decreases epoch times, and this causes the initial
decrease of epoch times in Figs. \ref{epoch_times}(a) and
\ref{epoch_times}(b). However, the ditch lineages are seeded by
mutants of fitness-$n$ individuals. For later epochs, fitness $n$
individuals have many more bits, $(n-1) K$, that need to be set to
specific configurations. They are, therefore, more likely to undergo
deleterious mutations. The proportion $P^n_n$ of fitness-$n$
individuals during epoch $n$ thus decreases as a function of $n$. This
tends to increase the epoch times since smaller $P^n_n$ implies a
lower rate of seeding of ditch lineages. Moreover, the probability
$\pi_n$ that a genotype of fitness $n+1$, once found, spreads through
the population decreases as $n$ increases as well. Eventually, these
effects start to dominate, and epoch times start rising again. In
fact, at a certain point, for large $n$, the probability $\pi_n$ may
become so small that fitness-$n+1$ individuals cannot be stabilized in
the population at all.  In this regime, the dynamics again reaches the
well known error threshold: the population dynamics cannot store the
necessary $nK$ bits of information that define epoch $n+1$.

In Figs. \ref{epoch_times}(c) and \ref{epoch_times}(d), there is no initial
decrease of epoch times: the epoch times increase monotonically as a function
of $n$. In this case, from the start of the runs the decrease of $\pi_n$ and
$P^n_n$ with $n$ dominates the effect of the relatively shallower
ditches. 


\begin{figure}[htbp]
\centerline{\epsfig{file=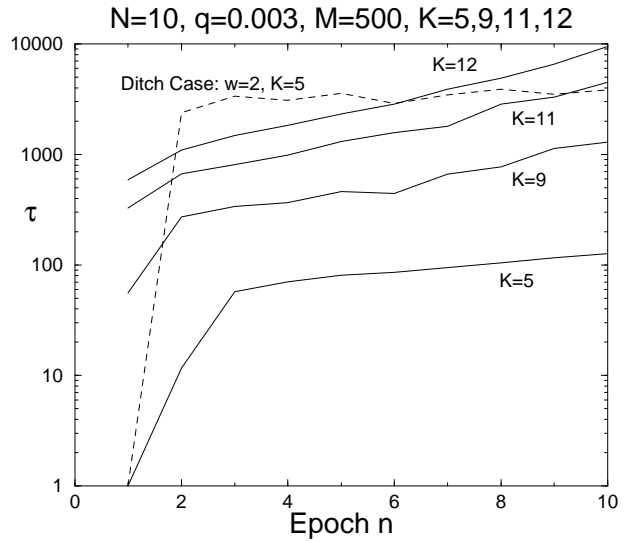,width=3.25in}}
\caption{Comparison of the entropy- and fitness-barrier crossing times 
for the Royal Staircase with Ditches fitness function. The dashed line
gives the epoch times for the case of {\em minimal width} ($w=2$)
ditches of height $h=1$. This data is the same as that in
Fig. \ref{epoch_times}(b).  The solid lines plot the entropy-barrier
epoch times for a fitness function without ditches ($w=0$) for several
different block lengths: $K=5$, $9$, $11$, and $12$. All other
parameters are identical to the ditch-case parameters: i.e., $N=10$
blocks, a mutation rate of $\mu = 0.003$, and a population size of
$M=500$. The vertical axis is shown on logarithmic scale.}
\label{comparison}
\end{figure}

Finally, we compare these epoch times, for ditches of height $h=1$ and
width $w=2$, with the epoch times for the entropy-barrier case ($w=0$)
at different block lengths $K$. This comparison is shown in
Fig. \ref{comparison}.  Epoch times $\tau$ are shown on a logarithmic
scale. The dashed line shows the experimentally obtained data for the
ditch case with $N=10$ blocks of length $K=5$, a (minimal) ditch width
of $w=2$, and height of $h=1$, a population size of $M=500$, and a
mutation rate of $\mu =0.003$; as used in
Fig. \ref{epoch_times}(b). The solid lines show experimentally
estimated epoch times for the neutral case ($w=0$), for several
different block lengths. All other parameters are the same.

In comparing the solid line for $K=5$ with the dashed line we see that
the introduction of this minimal ditch increases the epoch times by
{\em factors} ranging from $50$ to $250$. In order to obtain roughly
comparable epoch times for the neutral case $w=0$, one has to increase
the block length to as high as $K=12$. This demonstrates how much more
rapidly entropy barriers are crossed than fitness barriers. In the
time that it takes the population to cross a fitness barrier of width
$w=2$, the neutrally diffusing population will have crossed an entropy
barrier of an additional $12-5 = 7$ bits. Thus, the neutrally
diffusing population explores roughly $2^7 = 128$ times as many
different neutral configurations in the time that it takes to cross
the $2$-bit fitness barrier. As we will see below, the difference in
time scale for crossing fitness and entropy barriers grows much larger
for more realistic (lower) mutation rates and (longer) sequences.

\section{Conclusions}
\label{conclusions}

We analyzed in detail the barrier crossing dynamics of a population
evolving under selection and mutation in a constant selective
environment. Barriers of two distinct types exist: fitness barriers
and entropy barriers. Fitness barriers occur in a fitness
``landscape'' when genotypes of higher fitness than currently present
in the population are separated from the current most-fit genotypes by
valleys of lower fitness. In order for such barriers to be crossed, a
rare sequence of mutants must cross the valley of low-fitness
genotypes. The second type of barrier, entropy barriers, occurs when
genotypes of current best-fitness form large {\em neutral networks} in
genotype space that only have a small number of connections ({\em
portals}) to genotypes of higher fitness. Evolving populations diffuse
at random through these neutral subbasins under selection and
mutation. Since connections to higher fitness genotypes are rare, the
population must search and spread over large parts of the neutral
network, before higher-fitness genotype portals are discovered.

We will now qualitatively and quantitatively summarize the general
picture that has emerged from our analysis of barrier crossing
dynamics. The first important observation to be made is that there is
a large qualitative difference between the genealogies of individuals
of current best fitness and those with suboptimal fitness. All
suboptimal individuals in the population are relatively recent
descendants of genotypes with the current highest fitness. In other
words, individuals with suboptimal fitness only give rise to
genealogical bushes---genealogies of finite, typically short,
length. Individuals with the current highest fitness are the only ones
that give rise to lineages of potentially infinite length.

More formally, let us denote by $\Lambda$ the neutral network of
genotypes whose fitness equals that of the current highest-fitness
individuals. The subpopulation of the current population that is on
the neutral network $\Lambda$ then effectively acts as a source for
the whole population's descendants. Genealogies of lower-fitness
individuals outside of $\Lambda$ go extinct relatively quickly and are
replaced by new genealogies that are mutant descendants of individuals
on the neutral network $\Lambda$. Therefore, only lineages of
individuals on the network $\Lambda$ can ``travel'' long distances
through genotype space. The subpopulation of genotypes in $\Lambda$
diffuses randomly through $\Lambda$, eventually visiting almost all
genotypes in $\Lambda$ and all of their single mutant neighbors.

At the same time, the excess
reproduction of individuals in $\Lambda$, combined with deleterious
mutations, creates individuals in the lower-fitness valleys or ditches
around $\Lambda$. These individuals then give rise to genealogies of
valley individuals that, in turn, probe at random the genotype space
surrounding $\Lambda$. However, since these valley genealogies
typically go extinct quite rapidly, individuals in the valley never
travel far from the neutral network $\Lambda$. That is, only those
parts of the valleys that are in the immediate neighborhood of the
neutral network $\Lambda$ will be quickly explored. Valley genealogies
that cross a wide valley are very rare.

Therefore, on the shortest time scale, the population explores the
neutral network $\Lambda$ and its neighborhood, as shown in Fig.
\ref{Genealogies}. If there are any higher-fitness genotypes in the
immediate neighborhood of the network $\Lambda$, the population is
most likely to escape from the metastable state (epoch) by crossing
this entropy barrier. If the neutral network $\Lambda$ is completely
surrounded by valleys of lower fitness, the population will eventually
escape from the metastable state when a mutant crosses one of the
valleys that surrounds $\Lambda$, i.e., by crossing a fitness
barrier. Since the waiting time for such a fitness barrier crossing
increases very rapidly with the width of the barrier, it is most
likely that the escape will occur along one of the narrowest valleys
that connects $\Lambda$ to higher fitness genotypes. Although shallow
valleys are more easily crossed than deep valleys, the effect of
barrier height is relatively small compared to the effect of barrier
width.

This basic picture of the metastable population dynamics is captured
more quantitatively by the scaling relations of
Eqs. (\ref{analytic_scaling}) and (\ref{entropic_scaling}) for the
fitness-barrier and entropy-barrier crossing times, respectively. By
equating these, we get a rough comparison of the trade offs in the
crossing of entropy versus fitness barriers. To make the comparison
for more general cases, we replace the factor $2^L$ in the
entropy-barrier scaling form, Eq. (\ref{entropic_scaling}), with
$V_\Lambda$ to denote the volume of the neutral network
$\Lambda$. We also replace $L$ by the average {\em neutrality}
$\nu$---i.e., the average number of neutral neighbors---of genotypes
in $\Lambda$. This extends the entropy-barrier scaling form to
more general neutral network topologies than simple binary
hypercubes. We then obtain
\begin{equation}
V_\Lambda = \frac{\nu}{w!} \left(
\frac{\log(\sigma)}{\mu} \right)^{w-1} ~.
\label{FitnessVersusEntropyTime}
\end{equation}
This is an estimate of the volume $V_\Lambda$ of the network $\Lambda$
that is explored in the time it takes the population to cross a fitness
barrier of height $\sigma$ and width $w$.

Recall that the factor $w!$ denotes the number of
different paths leading across the fitness barrier and that
$\mu/\log(\sigma)$ gives the average number of mutations valley
lineages undergo before they go extinct. The relative size of $\mu$
and $\log(\sigma)$, together with the barrier width $w$, are the
decisive quantities. Only when $\mu/\log(\sigma)$ approaches $1$ can
fitness barriers be crossed relatively quickly. However, since $\mu$ is
typically very small, even in cases where the highest-fitness
genotypes have only a small fitness advantage over valley genotypes---i.e.,
$\sigma = 1 + \epsilon$---the determining ratio $\mu/\epsilon$ may
still be small.

As an illustration, assume that each genotype in $\Lambda$ has an average
of $\nu = 60$ neutral neighbors, that the mutation rate is $\mu = 10^{-6}$,
and that genotypes in $\Lambda$ are on average $1$ percent more fit than
genotypes in the valley. Substituting these values into Eq.
(\ref{FitnessVersusEntropyTime}) for a barrier of width $w=3$, we find
$V_\Lambda \approx 10^9$. Thus, before this fitness barrier will be
crossed, on the order of $10^9$ genotypes on the neutral network
$\Lambda$ will have been explored. Of course, these numbers should not
be taken as quantitative predictions, they are only rough order-of-magnitude
estimates. However, they do indicate the difference in time scales at
which entropy and fitness barriers are crossed.

In cases where the fitness advantage $\epsilon$ becomes so
small as to be comparable in size to $\mu$, the fitness barrier will
have effectively turned into an entropy barrier. That is, for such
small fitness advantages, selection is not able to stabilize the population
within $\Lambda$ and the population will freely diffuse through the
valleys. In short, before $\epsilon$ becomes so small as to be
comparable to $\mu$, the population will have already crossed the
error threshold. This can be seen, for instance, by taking the logarithm 
on both sides in Eq. (\ref{crit_sigma_infpop}).

As we discussed earlier, the picture of the barrier crossing dynamics
that has emerged from our analysis contrasts strongly with the common
view that this process is analogous to the dynamics of energy-barrier
crossing in physical systems. In those systems, the stochastic
dynamics follows the {\em local} gradient of the energy landscape.
This local character of the dynamics causes the barrier {\em height}
to be the main determinant of the barrier crossing time. Generally,
the evolutionary population dynamics cannot be described as
following a local gradient. The current highest fitness of the
individuals located on the neutral network $\Lambda$ sets an absolute
fitness scale against which valley individuals must
compete. Therefore, valley lineages are unlikely to survive for many
generations and so cannot undergo many mutations before going
extinct. This mechanism causes the barrier width to be the main
determinant of the fitness-barrier crossing time in population
dynamics.

The preceding results demonstrate the important role of {\em neutral
networks} in genotype space for evolutionary dynamics: where a
population goes in genotype space is largely determined by the
available neutral paths.

\section*{Acknowledgments}

The authors thank Paulien Hogeweg and Dan McShea for helpful comments
on the manuscript. This work was supported in part by NSF grant
IRI-9705830, Sandia National Laboratory, and the Keck Foundation. EvN's 
investigations were also supported by the Priority Program Nonlinear
Systems of the Netherlands Organization for Scientific Research (NWO).

\bibliography{epev} 
\bibliographystyle{plain}

\appendix

\section{Analytical Approximation of the Epoch Quasispecies}
\label{quasispecies_calc}

In this appendix, we construct the {\em generation operator} ${\bf G}$ 
for the class of Royal Staircase with Ditches fitness functions, and
analytically approximate the metastable quasispecies distributions
$\vec{P}^n$. 

As before, the operator ${\bf G}$ decomposes into a part ${\bf S}$
that encodes the expected effects of selection on the population
fitness distribution and a mutation operator ${\bf M}$ that encodes
the expected effects of mutation.

The selection operator $\bf S$ is easy to construct since selection
depends only on the population's fitness distribution. After
selection, we have for the nonditch genotypes that:
\begin{equation}
P^{\rm sel}_{n} \equiv \frac{\left( {\bf S} \cdot \vec{P}
 \right)_n}{\AvFit} = \frac{n}{\AvFit} P_n ~,
\label{RSwDitchSelectionOpI}
\end{equation}
and for the ditch genotypes that
\begin{equation}
P^{\rm sel}_{n,*} = \frac{n-h}{\AvFit} P_{n,*} ~.
\label{RSwDitchSelectionOpII}
\end{equation}

Construction of the mutation operator ${\bf M}$ is more involved, but
straightforwardly follows Ref. \cite{Nimw97b}. Formally, the probability
$\MuOp_{ij}$ that a genotype of type $j$ turns into a genotype of type
$i$ under mutation is given by a sum over all genotypes of type $i$ and
$j$, weighted by the probability of mutating from one genotype to the
other. When we refer to the ``type'' of a genotype we distinguish
genotypes only based on their location in an epoch's neutral network
($n$) or intervening ditch ($n,*$). We denote by $\Lambda_i$ the set of
all genotypes of type $i$ and by $d(s,s')$ the Hamming distance between
the genotypes $s$ and $s'$. Formally, we then have:
\begin{equation}
\MuOp_{ij} = \sum_{s \in \Lambda_i} \sum_{s' \in \Lambda_j}
\frac{\mu^{d(s,s')} (1-\mu)^{L-d(s,s')}}{| \Lambda_j |} ~,
\end{equation}
where $| \Lambda_j |$ is the size of the set $\Lambda_j$. That is, we
assume that a type $j$ genotype is equally likely to be any of its
possible $|\Lambda_j|$ genotypes. We then sum over all possible
genotypes of type $i$ to which $j$ can mutate. The generation operator
${\bf G}$ is then, as before, the product of the selection operator,
Eqs.  (\ref{RSwDitchSelectionOpI}) and (\ref{RSwDitchSelectionOpII}),
and the resulting mutation operator: ${\bf G} = {\bf M} \cdot {\bf
S}$.

We now list expressions for the different types of ${\bf G}$'s
components.  In order to simplify the formulas, we first define
\begin{equation}
\alpha = \left( \frac{\mu}{1-\mu} \right)^w ~,
\end{equation}
and the probability $\lambda$ to not mutate a block, which is given by
\begin{equation}
\lambda = (1-\mu)^K ~.
\end{equation}
Using this notation, the probability to mutate a block from type $A$
to type $B$, for instance, becomes
${\rm Pr}(A \rightarrow B) = \alpha \lambda$. Components $G_{ij}$
with $i<j$ become
\begin{equation}
G_{ij} = \alpha \lambda^{i-1} j ~,
\end{equation}
for $i \neq 1$, and
\begin{equation}
G_{1j} = \left( 1- \lambda^{j-1} -\alpha
\frac{\lambda-\lambda^{j-1}}{1-\lambda} \right) j ~,
\end{equation}
for $i = 1$. The diagonal components are given by
\begin{equation}
G_{jj} = \lambda^{j-1} j ~.
\end{equation}
For the ditch genotypes, the diagonal components are:
\begin{equation}
G_{(j-h)(j-h)} = \lambda^{j-1} (j-h) ~.
\end{equation}
Components describing the transitions from ditch genotypes to lower-fitness
nonditch genotypes $i < j$ are given by:
\begin{equation}
G_{i(j-h)} = \alpha \lambda^{i-1} (j-h) ~.
\end{equation}
Finally, mutations from nonditch genotypes to lower lying ditch genotype
come in three varieties. First, for $i < j-2$ we have 
\begin{equation}
G_{(i-h)j} = \alpha \lambda^{i-1} (1-\lambda-\lambda \alpha) j ~.
\end{equation}
For $i = j-1$ we have
\begin{equation}
G_{(j-1-h)j} = \lambda^{j-2} (1-\lambda-\alpha \lambda) j ~.
\end{equation}
And for the ditch lying immediately below, $i = j$, we have
\begin{equation}
G_{(j-h)j} = \frac{1}{2^{K}} \lambda^{j-2} (1-\lambda-\alpha \lambda) j ~.
\end{equation}

The quasispecies fitness distribution during epoch $n$ is given by the
principal eigenvector of the matrix ${\bf G}$ restricted to the components
with fitness lower than or equal to $n$; see Refs. \cite{Nimw97a} and
\cite{Nimw97b}. This eigenvector can be obtained numerically, using
the above formulas for the components $G_{ij}$.

An analytic approximation to $\vec{P}^n$ can be obtained by
considering mutations only from each fitness $j$ (or $j-h$) to equal
or lower fitness. This approximation is justified by the fact that
fitness-increasing mutations are very rare compared to fitness-decreasing
mutations. Under that approximation, the matrix ${\bf G}$
becomes upper triangular. For upper triangular matrices, the
components $\vec{P}^n_i$ of the principal eigenvector $\vec{P}^n$ obey
the equations:
\begin{equation}
\label{formal_qs_recurs}
P^n_i = \frac{\sum_{j>i}^n G_{ij} P^n_j}{G_{nn} -G_{ii}} ~,
\end{equation}
and, since $\vec{P}^n$ is a distribution, we additionally have the
normalization condition given by
\begin{equation}
\label{norm_cond}
\sum_{i=1}^n P^n_i = 1 ~.
\end{equation}
Note that the sums in the above equations involve both ditch genotypes
and nonditch genotypes. Their precise ordering with respect to fitness
depends, of course, on the barrier height $h$. For instance, for $0 <
h < 1$, the fitness $n-h$ genotypes in the ditch between fitness-$n$
and fitness-$(n+1)$ genotypes lie between fitness $n-1$ and $n$.

With the analytic expressions for the components of ${\bf G}$ in hand,
and by using Eq. (\ref{formal_qs_recurs}), we can express all $P^n_i$
for $i<n$ as a function of $P^n_n$. For instance, assuming $h < 1$,
such that the genotypes of fitness $n-h$ are the second
highest-fitness genotypes in the population, we have
\begin{equation}
P^n_{n,*} = \frac{n (1-\lambda -\alpha \lambda)}{2^K h \lambda} P^n_n ~.
\end{equation}
Using this, we have for genotypes of fitness $n-1$ that
\begin{equation}
P^n_{n-1} = \frac{n \left( \alpha (n-h)(1-\lambda-\alpha \lambda)+2^K
\lambda \right)}{\left( 1-n(1-\lambda) \right) \lambda 2^K} P^n_n ~.
\end{equation}
When all components $P^n_i$ with $i<n$ have been expressed in terms of
$P^n_n$ in this way, one finally uses the normalization condition
Eq. (\ref{norm_cond}) to determine $P^n_n$.

This procedure leads in a straightforward way to a relatively accurate 
analytical expression for the epoch fitness distributions. The
expressions, however, are rather cumbersome and we will not list them
all here. In general, they depend on the size of $h$ since the ordering of
the fitness values of the different types of genotypes depends on $h$.

\end{multicols}

\end{document}